\documentclass[prb,preprint,floatfix]{revtex4-2}
\usepackage{amsmath,graphicx,graphics,subfigure,epsfig,hyperref}

\usepackage{verbatim, color}

\newcommand{\be}{\begin{equation}}
\newcommand{\ee}{\end{equation}}
\newcommand{\ba}{\begin{eqnarray}}
\newcommand{\ea}{\end{eqnarray}}

\begin{document}

\setcitestyle{super}
\title{Smooth double barriers in quantum mechanics}
\author{Avik Dutt}
\email{quantumavik@gmail.com}
\affiliation{Department of Electronics and Electrical Communication
Engineering, Indian Institute of Technology, Kharagpur, 721302, India}

\author{Sayan Kar}
\email{sayan@iitkgp.ac.in}
\affiliation{Department of Physics and Meteorology and Center for Theoretical Studies,
Indian Institute of Technology, Kharagpur, 721302, India}

\begin{abstract}
Quantum mechanical tunneling across smooth double barrier potentials 
modeled using Gaussian functions, is analyzed numerically and by using the 
WKB approximation.
The transmission probability, resonances as a function of incident particle energy, and
their dependence on the barrier parameters are obtained for various
cases. We also discuss the tunneling time, for which we obtain generalizations of
the known results for rectangular barriers.
\end{abstract}

\maketitle

\section{Introduction}

The usual example of quantum mechanical tunneling is the rectangular barrier in one dimension.\cite{beiser, cohen, merzbacher} Curious students might wonder what happens if we consider smooth barriers. Do the tunneling results remain the same? Are there quantitative or
qualitative differences? Students might also wonder what
happens if there is more than one barrier.
To answer these questions (particularly for smooth barriers),
numerical methods and approximations are essential because little can be done analytically. In this paper we discuss
tunneling across smooth double barriers of various types.

Double barrier potentials arise
in many diverse areas. We give a few examples in the following.

{\em Quantum heterostructures}.\cite{heterostructures,heterostructures2}
Semiconductor heterostructures are layered, thin (about 100 nanometers or less)
sandwich structures made with different semiconductor materials (for example, GaAs between two AlAs layers). The existence of junctions of different materials
is the reason for the occurrence of sequences of wells and barriers in such structures. Semiconductor heterostructures have
made it possible to control the parameters inside crystals and devices such as
band gaps and the effective masses of charge carriers and their mobilities, refractive indices, and electron energy spectrum. Heterostructure electronics
are widely used in many areas, including laser-based telecommunication
systems, light-emitting diodes, bipolar transistors, and low-noise high-electron-mobility transistors for high-
frequency applications including satellite
television.\cite{heterostructures2}

{\em High energy physics}.
A barrier penetration model has been developed
for heavy ion fusion,\cite{pddbphh} taking into account the realistic features of the Coulomb potential for the case of nuclei. The typical barrier shapes encountered for nuclei conform to double barriers.

{\em Extra dimensions}.
Another area where such potentials arise is in the physics of extra dimensions. In particular,
they appear in the context of localization of particles and fields
on a four-dimensional hypersurface, known as the brane.
In these brane-world models\cite{rs} localized particles in
four dimensions are viewed as bound or quasi-bound (resonant) states in a double barrier effective potential spread across the extra dimensional
coordinate.

{\em Nonlinear Schr\"{o}dinger equation}.
The nonlinear Schr\"{o}dinger equation has been used 
in investigations of Bose-Einstein condensates
to probe macroscopic quantum tunneling\cite{rapedius}$^{-}$\cite{paul} and 
gravity surface waves in fluids, among many other applications. The methods we will discuss here may be extended, with minor modifications, to the Gross-Pitaevskii equation, 
a nonlinear Schr\"{o}dinger equation arising in the study of many body systems. Above
barrier reflection and tunneling in the nonlinear Schr\"{o}dinger equation is discussed in Ref.~[9]. 
The double Gaussian barrier, as well as the rectangular barrier, has been used as model potentials in studies\cite{rapedius,paul} of the nonlinear Schr\"{o}dinger equation. However,
the definition of the tunneling coefficient is ambiguous for weak nonlinearity, because the
principle of superposition does not strictly hold.

Apart from the applications we have discussed, we mention some recent 
pedagogical articles on double barrier tunneling. 
Numerical solutions for quantum tunneling through rectangular double barrier
heterostructures have been explored.\cite{mendez} Further, 
a detailed analysis of the propagation of wave packets across double barriers has been
performed.\cite{double} Three-dimensional \textcolor{black}{$s$}-wave tunneling for double barrier potentials has also 
been investigated.\cite{pramana}

The paper is organized as follows. In Sec.~II we introduce the
various types of double barriers that will be discussed.
Section~III discusses the theory of double barrier tunneling,
and the various methods we will use. Numerical and
semiclassical analyses of the smooth barriers mentioned in Sec.~II
are carried out in Sec.~IV. In Sec.~V we determine the tunneling time across
smooth double barriers. Section~VI summarizes our results
and mentions some possibilities for future studies.

\section{Tunneling across double barriers}

A simple way to model double barrier potentials is to add a term shifted in position to the single barrier
potential function, that is,
\begin{equation}
\label{eq:eqn1}
V(x) = V_{\rm single}(x,V_1,w_1) + V_{\rm single}(x-a,V_2,w_2),
\end{equation}
where $V_{1,2}$ and $w_{1,2}$ are the height and width of the two barriers. 
The separation between the barrier heights is $a$. 

\begin{figure}[h!]
\centering
\includegraphics[width=4in]{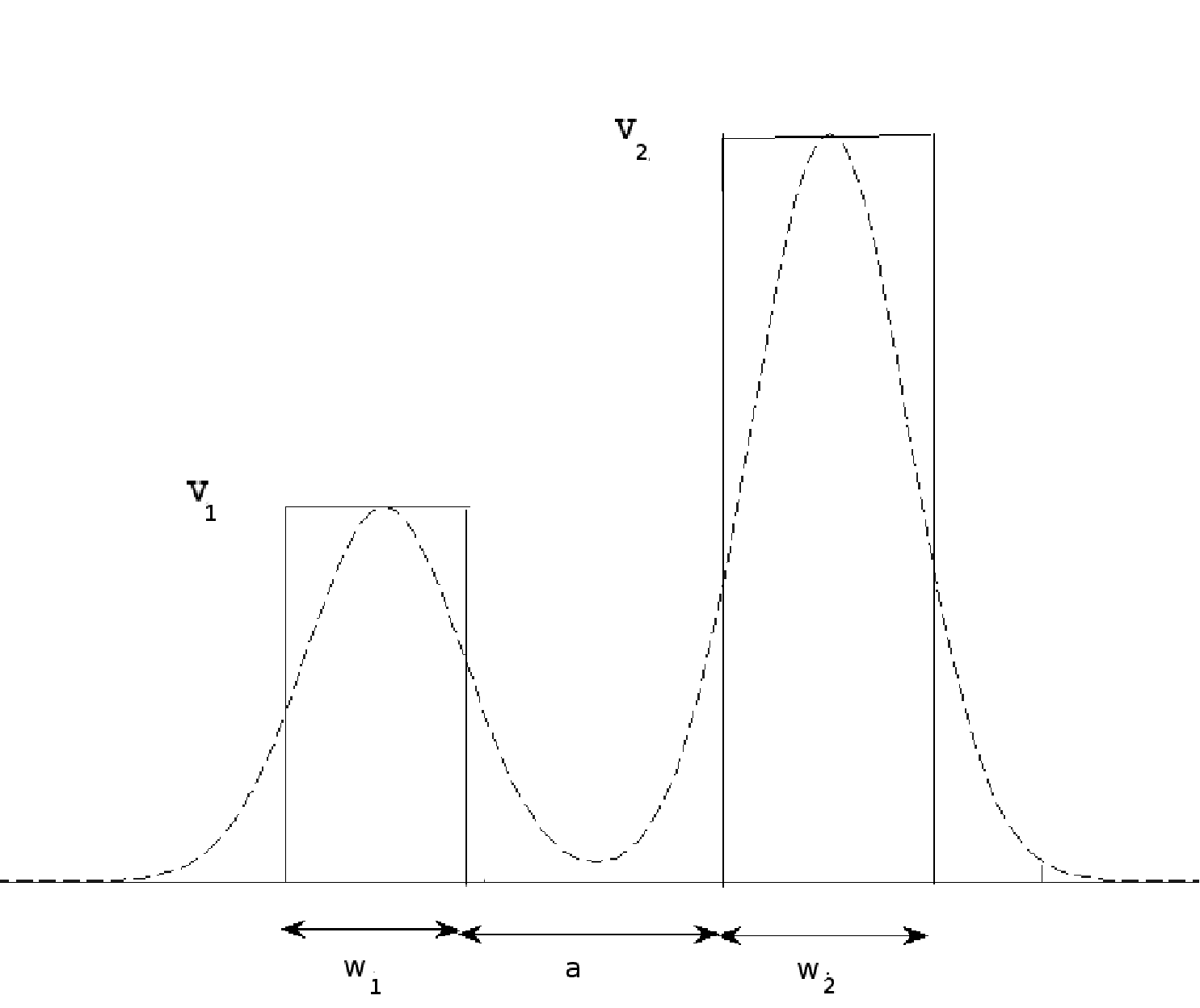}
\caption{Rectangular double barrier (solid line) of widths $w_1$ and $w_2$ separated by the distance $a$. The heights of the two single barrier potentials are $V_1$ and $V_2$. Also shown is a double Gaussian barrier (dashed line).}\label{fig1}
\end{figure}

The simplest and the easiest barrier to analyze is the rectangular double potential barrier, whose parameters are defined in Fig.~\ref{fig1}. We have
\begin{equation}
\label{eq:eqnsquare}
V_{\rm single}(x,V_1,w_1) = \begin{cases}
0 & x<0\\
V_1 & 0\leq x \leq w_1\\
0 & x>w_1
\end{cases}.
\end{equation}

Similarly, the Gaussian double barrier (see Fig.~\ref{fig1}) is written 
using 
\begin{equation}
\label{eq:eqngauss}
V_{\rm single}(x,V_1,\sigma_1) = V_1 e^{-x^2/2\sigma_1^2}
\end{equation}
and hence $V(x) = V_{\rm single}(x,V_1,\sigma_1)+V_{\rm single}(x-a,V_2,\sigma_2)$.

The time-independent Schr\"{o}dinger equation for a one-dimensional potential 
$V(x)$ is
\begin{equation}
\label{eq:Schrodinger}
\frac{d^{2} \psi }{dx^{2}} + \frac{2m}{\hbar^{2}}[E-V(x)] \psi = 0,
\end{equation}
where $m$ is the mass of the particle, $E$ is its energy, and $\psi$ is the 
wavefunction.
The Schr\"{o}dinger equation for a rectangular double barrier can be readily 
solved for $\psi$, giving five solutions in the five regions.
The solutions and their derivatives can be matched at the boundaries
between two regions to obtain a relation between the incoming and outgoing waves. The transmission amplitude is\cite{doublebarrier}
\begin{eqnarray}
\label{eq:sqtrans}
T & =& 4[ e^{ik_{1}(w_1+w_2)}(2\cosh(k_{3} w_2) - ik_{1-3}\sinh(k_{3} w_2)) \textcolor{black}{\times (2\cosh(k_{2} w_1) - ik_{1-2}\sinh(k_{2} w_1))} \nonumber \\
&&{} + e^{ik_{1}(g+b-w_1)}k_{1+3}k_{1+2}\sinh(k_{2} w_1)\sinh(k_{3} w_2)]^{-1},
\end{eqnarray}
where $w_1$ and $w_2$ are given in Eq.~\eqref{eq:eqn1},\eqref{eq:eqnsquare}. The other parameters are defined as
\begin{subequations}
\begin{align}
g & =w_1+w_2, && b=w_1+a+w_2 \\
k_{1} & =\sqrt{2mE}/\hbar, && k_{2} =\sqrt{2m(V_{1}-E)}/\hbar \\
k_{3} & =\sqrt{2m(V_{2}-E)}/\hbar), && k_{i \pm j} = \dfrac{k_{i}}{k_{j}} \pm \dfrac{k_{j}}{k_{i}}, \quad (i,j = 1,2,3).
\end{align}
\end{subequations}
The transmission coefficient (probability) is $ \mathcal{T} = T^{*}T$.

For the rectangular double barrier, for which we have analytical solutions, we match $\psi$ and $d\psi/dx$ at the discontinuities to obtain the relations between the wavefunction amplitudes, that is, $A$ and $C$. For a smooth barrier, the ratio of $A$ and $C$ can be obtained by solving Eq.~(\ref{eq:Schrodinger}) numerically.
The ratio of the maxima on the right-hand side of the barrier to the incident amplitude on the left-hand side of the barrier can be found from the plot of $\psi$ versus $x$.

The analysis can be extended to the arbitrary potential functions.\cite{riccoazbel,cohen}
However, exact analytical solutions of Eq.~(\ref{eq:Schrodinger}) are usually difficult or impossible to find. In particular, it is impossible to exactly solve the Schr{\"o}dinger equation for the Gaussian potential.
To obtain useful results, we either find numerical solutions or use the WKB approximation.

To solve the Schr\"{o}dinger equation numerically, we write Eq.~(\ref{eq:Schrodinger})
as a system of two first-order linear ordinary differential equations with
$\phi_{1} = \psi$ and
$\phi_{2} = d\psi/dx$.
With these substitutions, Eq.~(\ref{eq:Schrodinger}) becomes
\begin{subequations}
\label{eq:numerical}
\begin{align}
\phi _{1}' & = \phi_{2},\\
\phi _{2}' & = \frac{2m}{\hbar^{2}}(V(x)-E) \phi_{1}.
\end{align}
\end{subequations}
Equation~\eqref{eq:numerical} can be solved for $\phi_{1}$ to obtain $\psi(x)$.
We used the fourth order Runge-Kutta algorithm.

We compute the wavefunction on both sides of the barrier for different energies in the required range (up to twice the maximum potential height of the barriers). Because the potentials are rapidly decaying and we are mainly interested in the barrier penetration properties, the use of asymptotic solutions for $\psi$ to calculate $\mathcal{T}$ is justified.
To the far right and far left of the well, the forms of $\psi$ are,
\begin{equation}
\psi_{\rm right} \rightarrow C e^{ikx} , \quad \psi_{\rm left} \rightarrow A e^{ikx} + B e^{-ikx}.
\end{equation}
The transmission coefficient is calculated from
\begin{equation} \mathcal{T} = \frac{|C|^{2} }{ |A|^{2}}.
\end{equation}

The Wentzel-Kramers-Brillouin (WKB) approximation\cite{powellcrasemann, merzbacher} can be used to obtain partially analytic results for an arbitrary potential well. In this approximation the wavefunction $\psi$ is expressed as a power series in $\hbar$. We write the wavefunction in the form
\begin{equation}
\psi = A(x) e^{i \phi(x)}.
\end{equation}
If we assume the potential is slowly varying and neglect the second derivatives of $\phi$ and $A$, the solutions of the Schr\"{o}dinger equation can be expressed as
\begin{equation}
\label{eq:wkbpsi}
\psi = \tilde A |p|^{-1/2} e^{\pm i \int |p|\, dx/\hbar}, \quad p=\sqrt{2m[E-V(x)]}.
\end{equation}
The limits of integration in Eq.~\eqref{eq:wkbpsi} are determined by the classical turning points at which $E=V(x)$. At these points the semiclassical approximation fails because the wavefunction becomes infinite. To the far left and to the far right of the barriers, the WKB method gives reasonably good approximations for $\psi(x)$. The matching of $\psi$ near the classical turning points is done by the use of special functions. The asymptotic forms of these functions result in a net phase shift of $\pi /4$ when going from a classically forbidden region to a classically allowed region or vice-versa.\cite{powellcrasemann, merzbacher}

\begin{figure}[h!]
\centering
\includegraphics[width=0.5\textwidth]{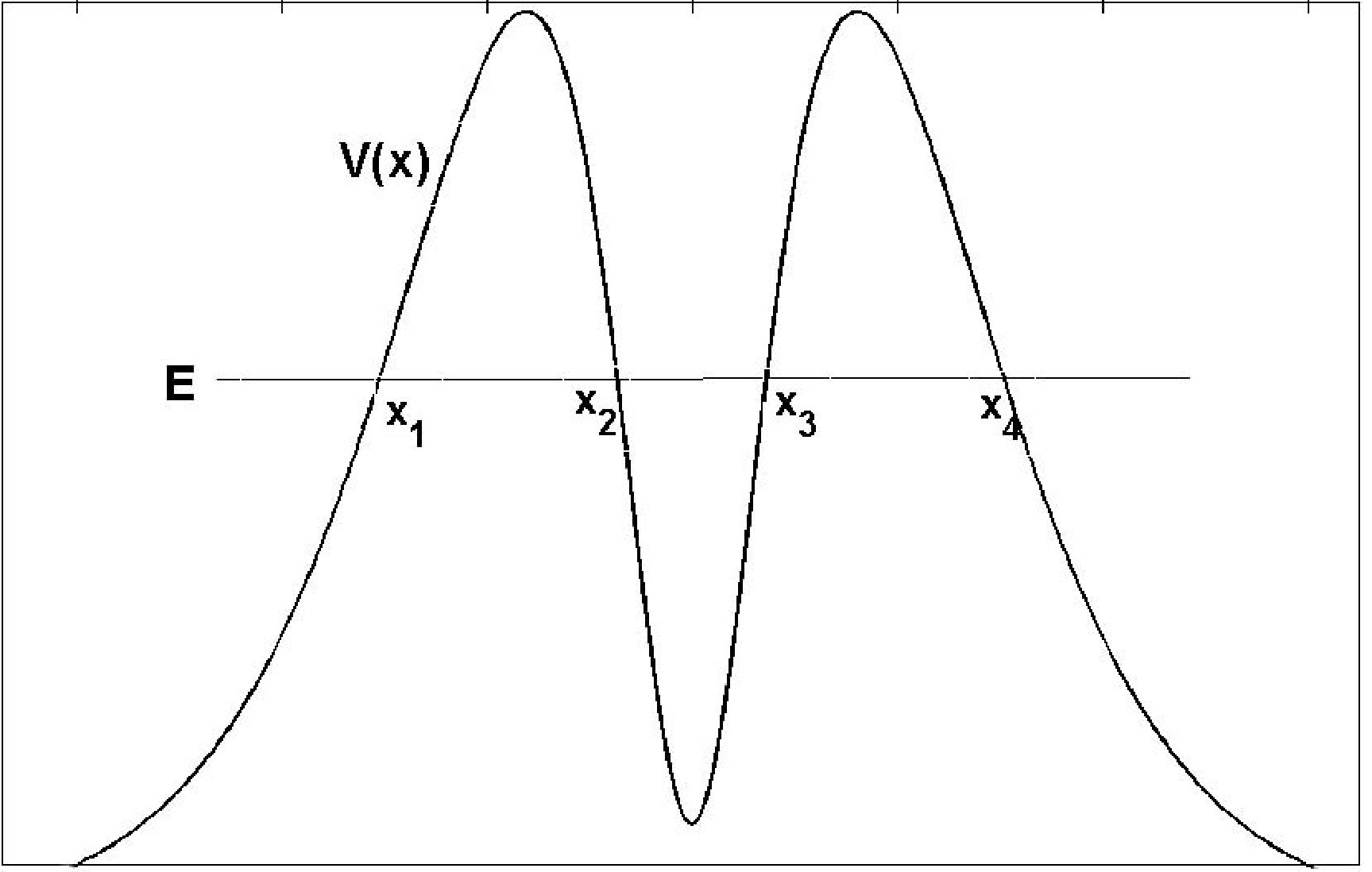}
\caption{An arbitrary double barrier potential $V(x)$ with classical turning points at $x_1$, $x_2$, $x_3$, and $x_4$. The regions $x_1< x < x_2$ and $x_3<x<x_4$ are classically forbidden, but the particle can tunnel through these regions quantum mechanically.}
\label{fig:wkb}
\end{figure}

For a general single barrier potential,
the transmission probability is\cite{powellcrasemann}
\begin{align}
\mathcal{T} & \approx \left(\frac{1}{T} + \frac{T}{4}\right)^{-2} = T^{2}/(1 + T^{2}/4)^{2}, \\
\noalign{\noindent where}
T &=\exp \left[-\int_{x_1}^{x_2} \sqrt{2m(V(x)-E)} \,dx /\hbar \right],
\end{align}
and $x_1$ and $x_2$ are the classical turning points.
The WKB approximation introduces errors which are consistent with $T$ being small for the single barrier case. Thus $\mathcal{T} \simeq T^{2}$, or
\begin{equation}
\mathcal{T} \simeq \exp\left[-\frac{2}{\hbar}\int_{x_{1}}^{x_{2}} \sqrt{2m(V(x)-E)} \,dx\right].
\end{equation}

For a double barrier we obtain four classical turning points instead of two as evident from Fig.~\ref{fig:wkb}.
Starting with the wave function on the far right of the barrier (region V), we have
\begin{equation}
\psi_{V} = A p^{-1/2} \exp\left(i \!\int_{x_{4}}^{x} p/\hbar \, dx + i\frac{\pi}{4}\right). \\
\end{equation}
Let $T_1$, $T_2$, and $T_3$ be defined as
\begin{subequations}
\begin{align}
T_1 & = \exp \left(-\int_{x_3}^{x_4} |p|/\hbar \, dx \right) \\
T_{2} & = \int_{x_{2}}^{x_{3}} p/\hbar \, dx \\
T_{3} & = \exp (- \int_{x_{1}}^{x_{2}} |p|/\hbar \, dx)
\end{align}
\end{subequations}
We expand the complex exponentials using Euler's identity, and then match the real and imaginary parts of $\psi$ near $x<x_{4}$ and $x>x_{4}$, and then near $x<x_3$ and $x>x_3$ (by applying the connecting formulae of the WKB method) and obtain
\begin{eqnarray}
\psi_{\rm III} & =& iA p^{-1/2} \Bigg[\Big(\frac{T_{1}}{4} -\frac{1}{T_{1}} \Big) \exp \Big(i\!\int_{x}^{x_{3}} p/\hbar \, dx + \frac{i\pi}{4}\Big) \nonumber \\
&&{}+ \Big(\frac{T_{1}}{4} + \frac{1}{T_{1}}\Big)\exp \Big(-i\!\int_{x}^{x_{3}} p/\hbar \, dx - \frac{i\pi}{4}\Big) \Bigg].
\end{eqnarray}
Similarly, we find
\begin{eqnarray}
\psi_I &=& \frac{A}{\sqrt{|p|}}\Bigg[\Big(\frac{C_{3} + C_{4}}{iT_{3}} + \frac{iT_{3}C_{4} - iT_{3}C_{3}}{4}\Big)\exp\Big(i\!\int_{x}^{x_{1}} p/\hbar \, dx + \frac{i\pi}{4}\Big) \nonumber \\
&&{} + \Big(-\frac{C_{3} + C_{4}}{iT_{3}} + \frac{iT_{3}C_{4} - iT_{3}C_{3}}{4} \Big)\exp \Big(-i\!\int_{x}^{x_{1}} p/\hbar \, dx - \frac{i\pi}{4} \Big) \Bigg].
\end{eqnarray}
The second term can be identified with the wave incident from the left, and yields the transmission coefficient
\begin{equation}
\label{eq:transmission}
\mathcal{T} \approx \left |\frac{\psi_{V}}{\psi_{\rm incident}} \right |^{2} = \left|\frac{T_{3}(C_{4} - C_{3})}{4} + \frac{C_{3}+C_{4}}{T_{3}} \right|^{-2}
\end{equation}
where
\begin{subequations}
\begin{align}
C_{3} & = (1/T_{1} - T_{1}/4)\exp(iT_{2})\\
C_{4} & = (T_{1}/4 + 1/T_{1})\exp(-iT_{2})
\end{align}
\end{subequations}
and $p(x) = \sqrt{2m[E-V(x)]}$.

Equation~(\ref{eq:transmission}) can be applied to a Gaussian double barrier with the assumptions that $E<V_{1},V_{2}$, $a\gg \sigma_{1},\sigma_{2}$, and $a-3\sigma_{1}- 3\sigma_{2} > 0$.
Hence, $x_{1,2} = \pm \sigma_{1}\sqrt{2(\log V_{1} - \log E)}$, and $x_{3,4} = a \pm \sigma_{2}\sqrt{2(\log V_{2} - \log E)}$. The integrations in the WKB method were performed using Simpson's rule.

\section{Numerical and semiclassical analysis for model smooth barriers}

\subsection{Numerical Analysis}

\begin{figure}[h!]
\centering
 \subfigure{\includegraphics[width=3in,height=1.7in]{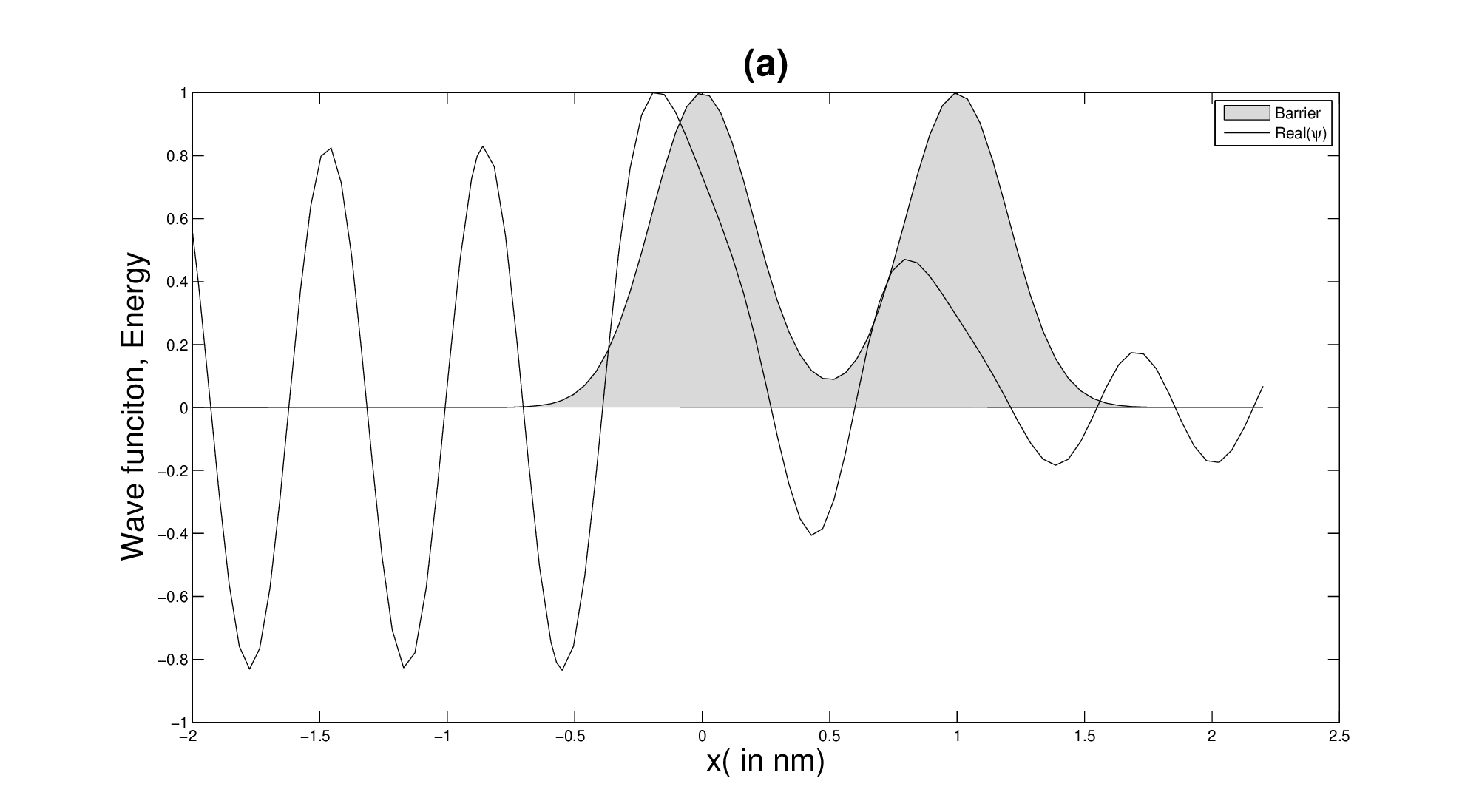}
 \includegraphics[width=3in,height=1.7in]{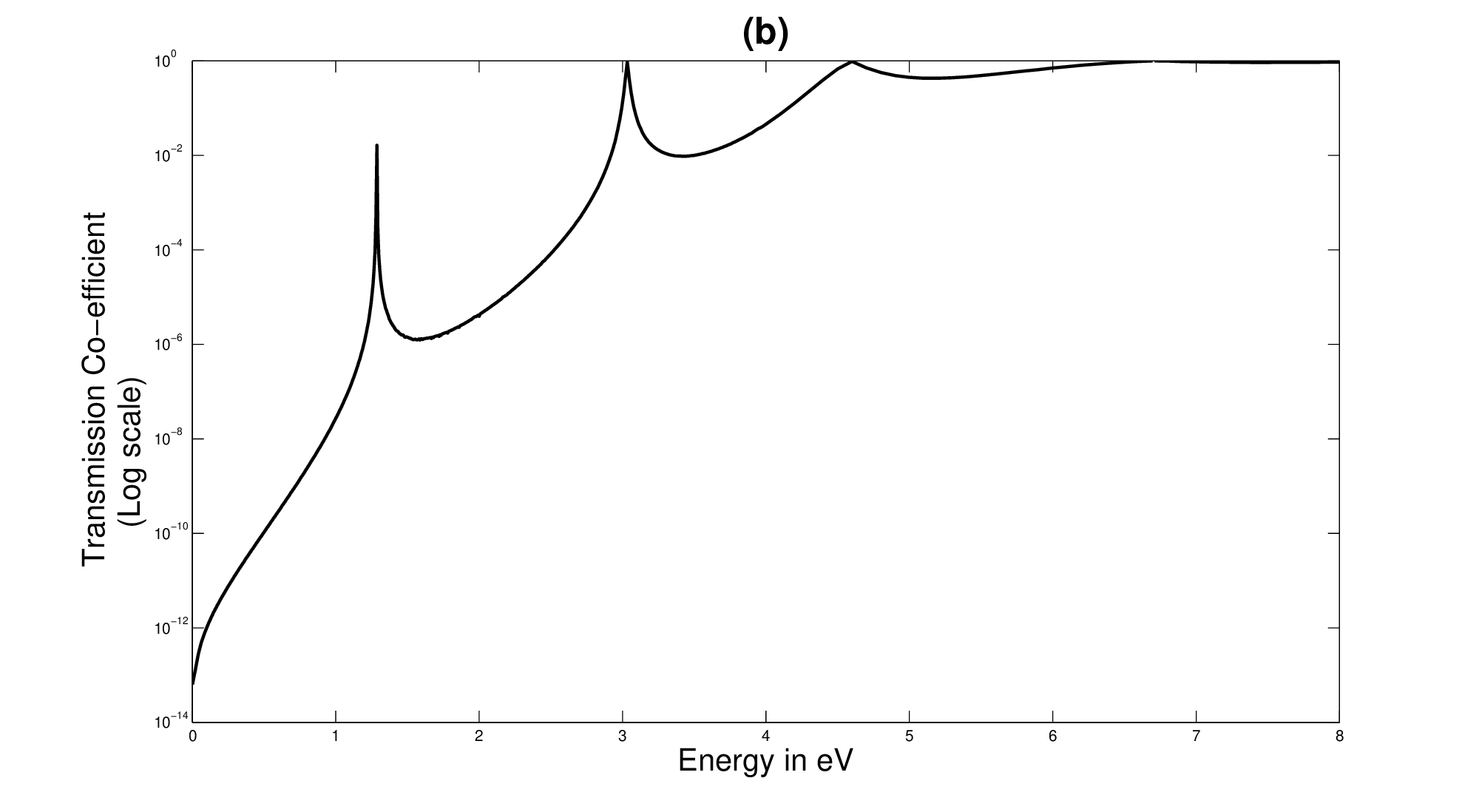}}
 \subfigure{\includegraphics[width=3in,height=1.7in]{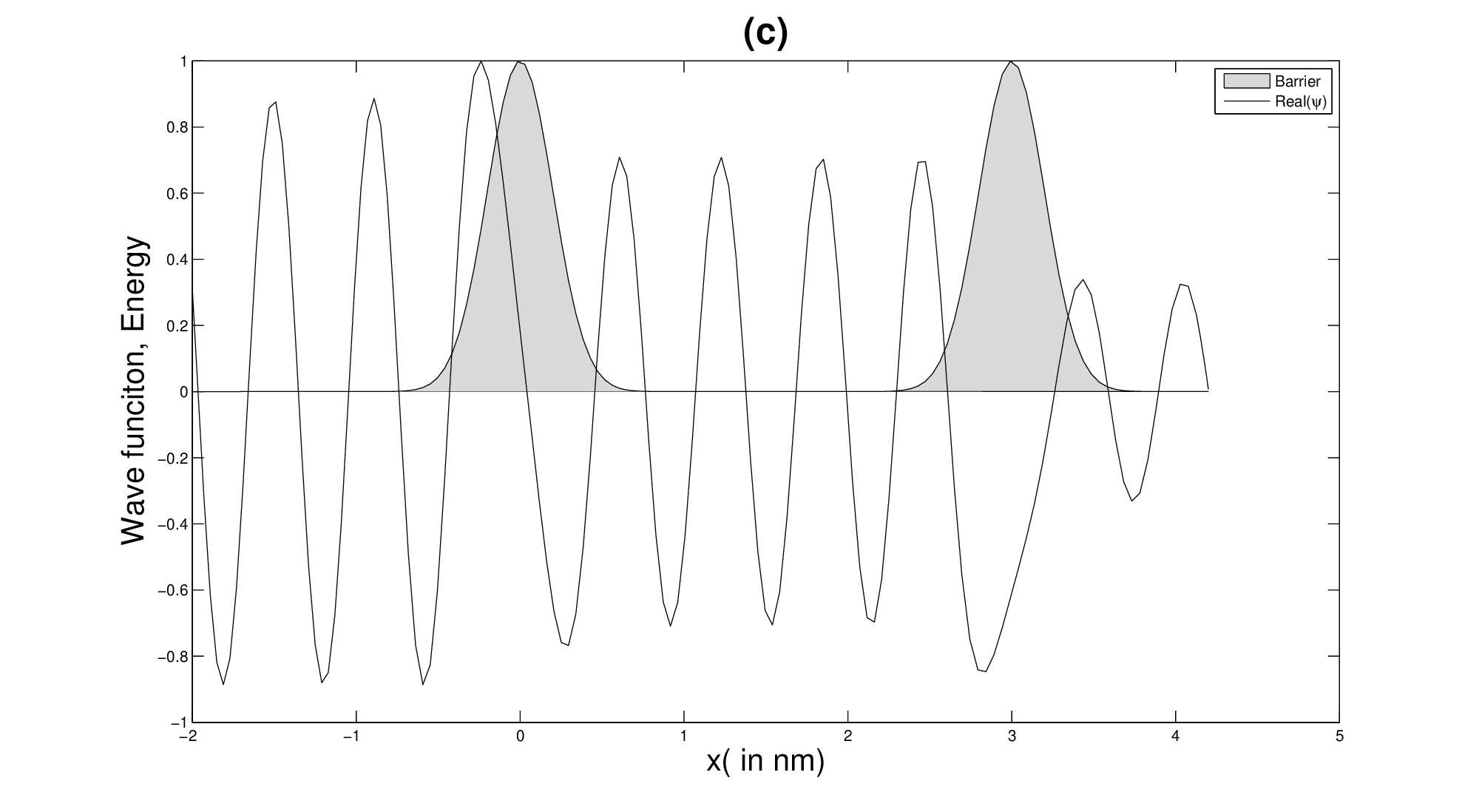}
 \includegraphics[width=3in,height=1.7in]{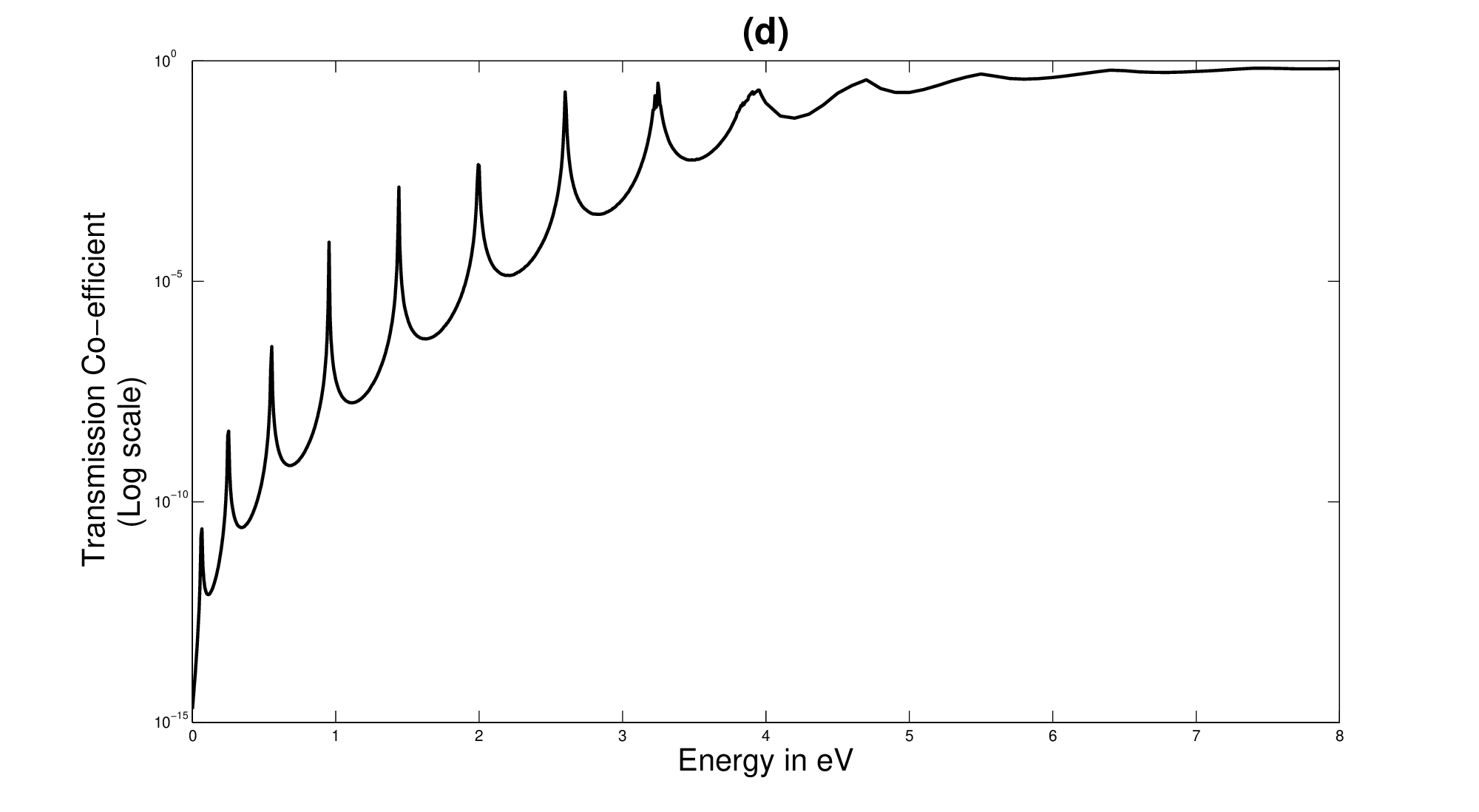}}
 \subfigure{\includegraphics[width=3in,height=1.7in]{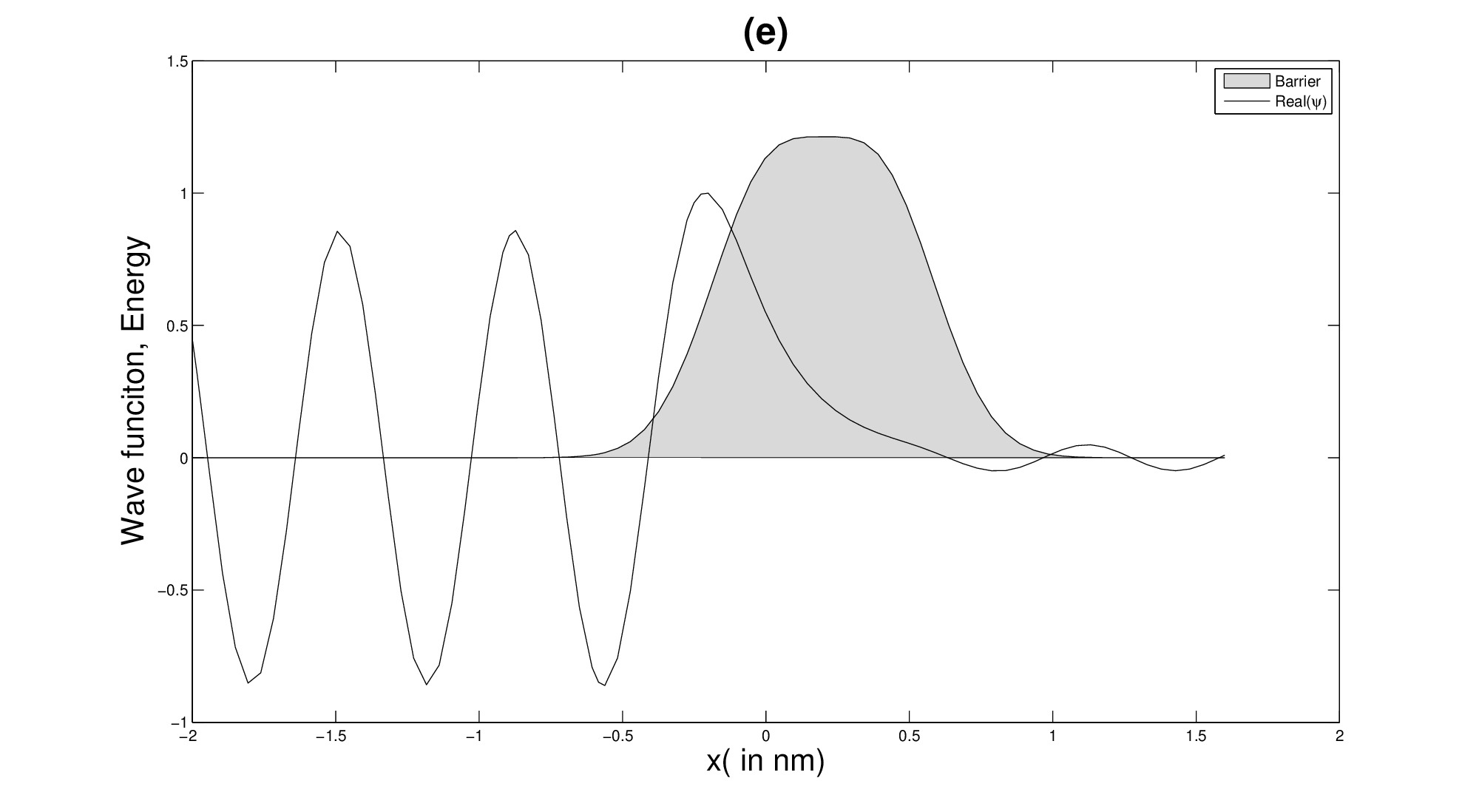}
 \includegraphics[width=3in,height=1.7in]{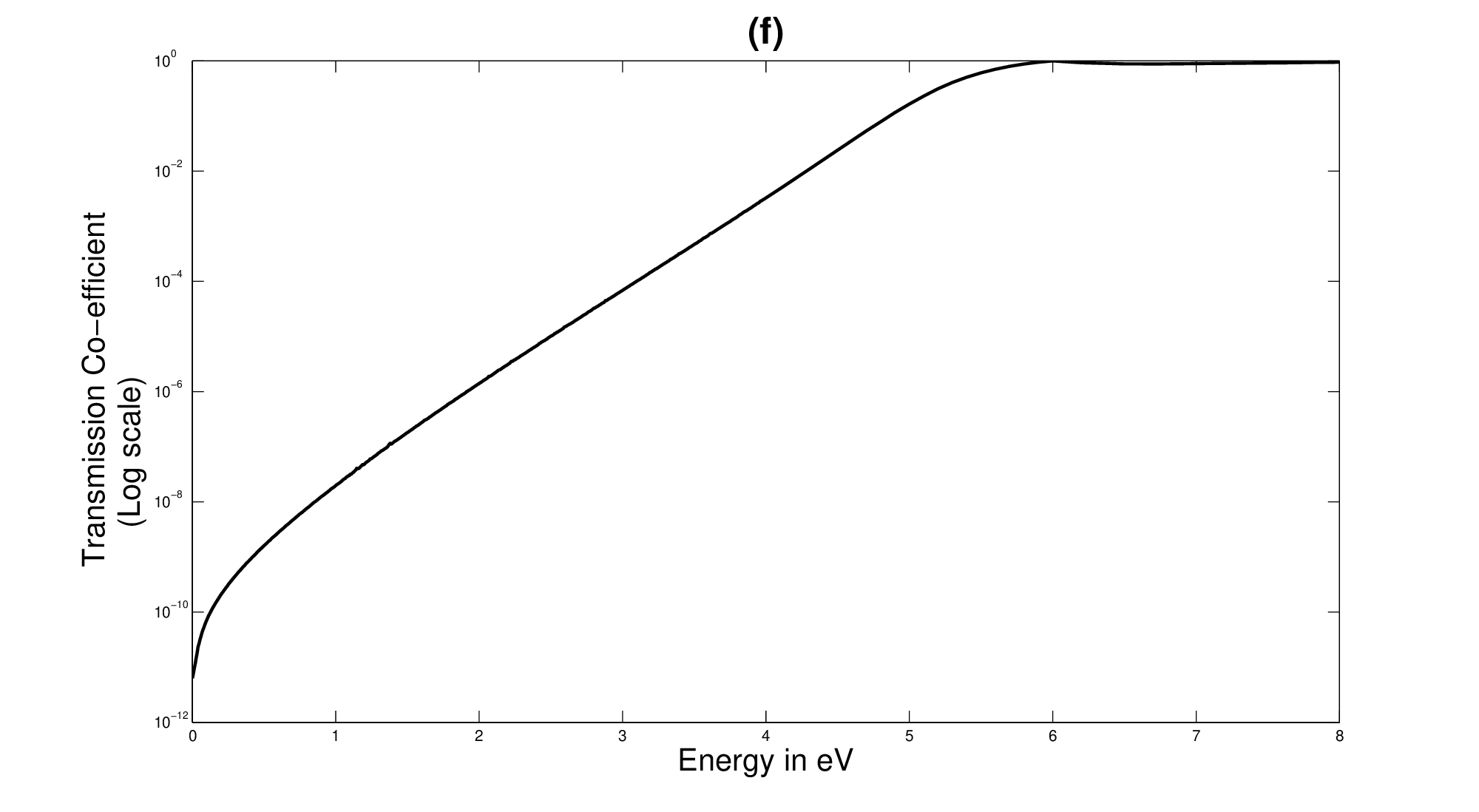}}\clearpage
 \subfigure{\includegraphics[width=3in,height=1.7in]{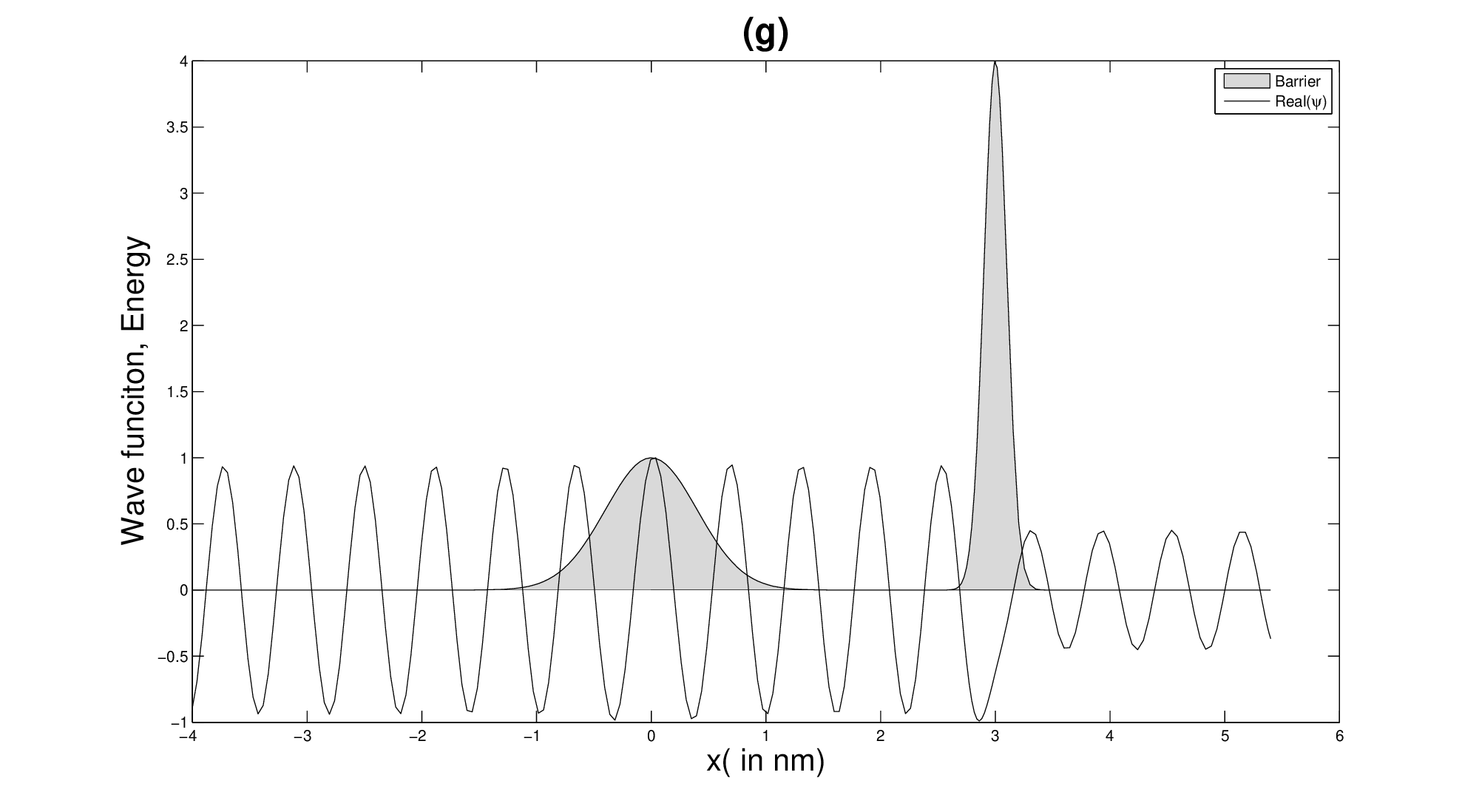}
 \includegraphics[width=3in,height=1.7in]{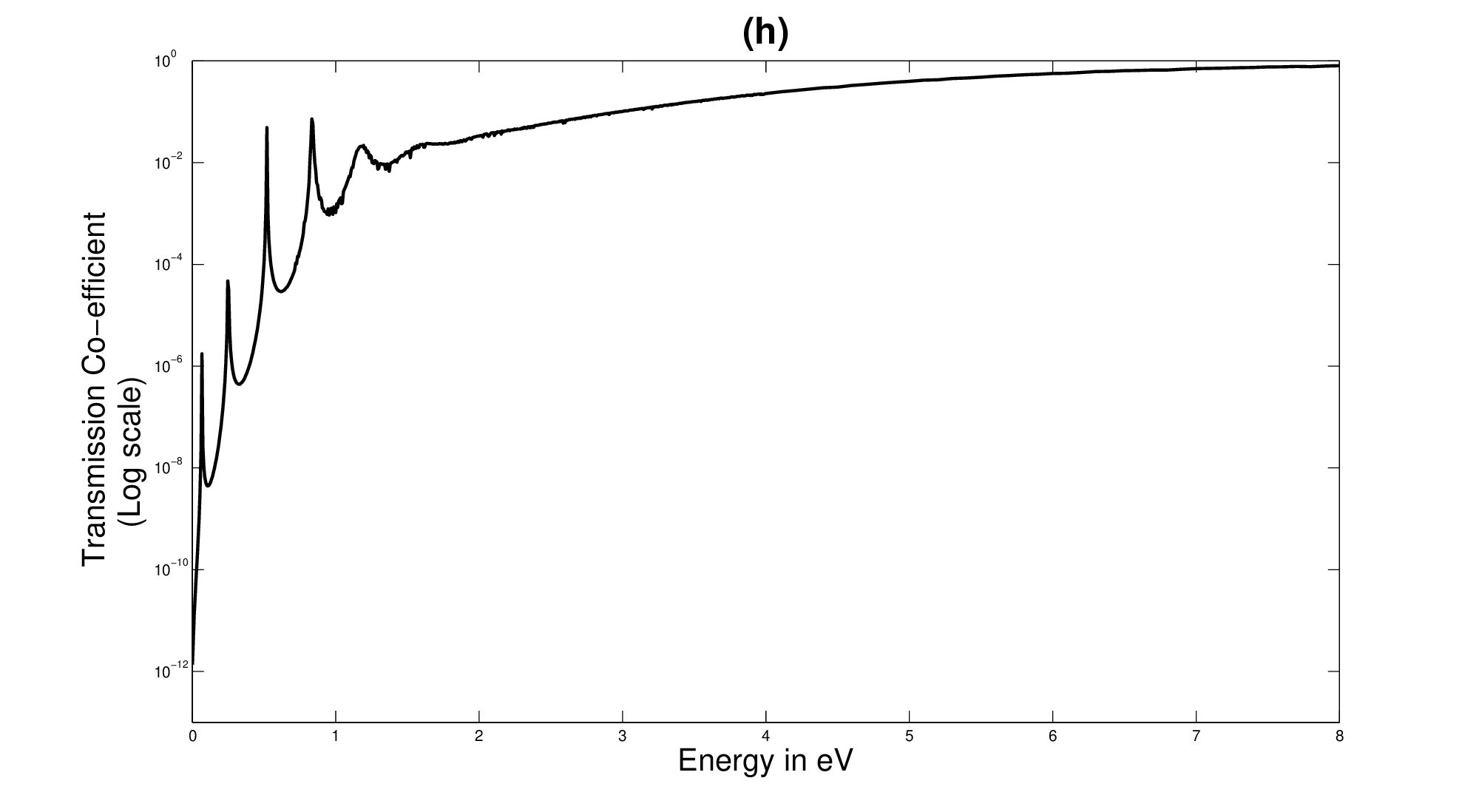}}
\vspace{-.2in}
\caption{{\footnotesize Plot of the potential (shaded) and wave function. The plots of the wave function are for an incident
energy of 4 eV. 
(a) $V_1=V_2=4$\,eV, $\sigma_1=\sigma_2=0.2$\,nm, $a=1$\,nm.
(b) Transmission probability for the barrier in (a).
(c) $V_1=V_2=4$\,eV, $\sigma_1=\sigma_2=0.2$\,nm, $a$ increased to 3\,nm.
(d) Transmission probability for the barrier in (c) -- the number of 
resonances is greater due to the larger
separation-to-width ratio. 
(e) $V_1=V_2=4$\,eV, $\sigma_1=\sigma_2=0.2$\,nm, $a$ reduced to 0.4\,nm so that the two Gaussian potentials almost merge into a single barrier.
(f) Transmission probability for the barrier in (e)-- the plot resembles the 
transmission probability of a single Gaussian barrier. 
(g) $V_1=1,\, V_2=4$\,eV, $\sigma_1=0.4$, $\sigma_2=0.1$\,nm, $a=3$\,nm.
(h) Transmission probability for the barrier in (g)--
asymmetrical barriers have a marked effect on the transmission probability.
}}
\label{fig:numerical}
\end{figure}

\begin{figure}[!h]
\centering
\subfigure{\includegraphics[width=3.2in,,height=2.2in]{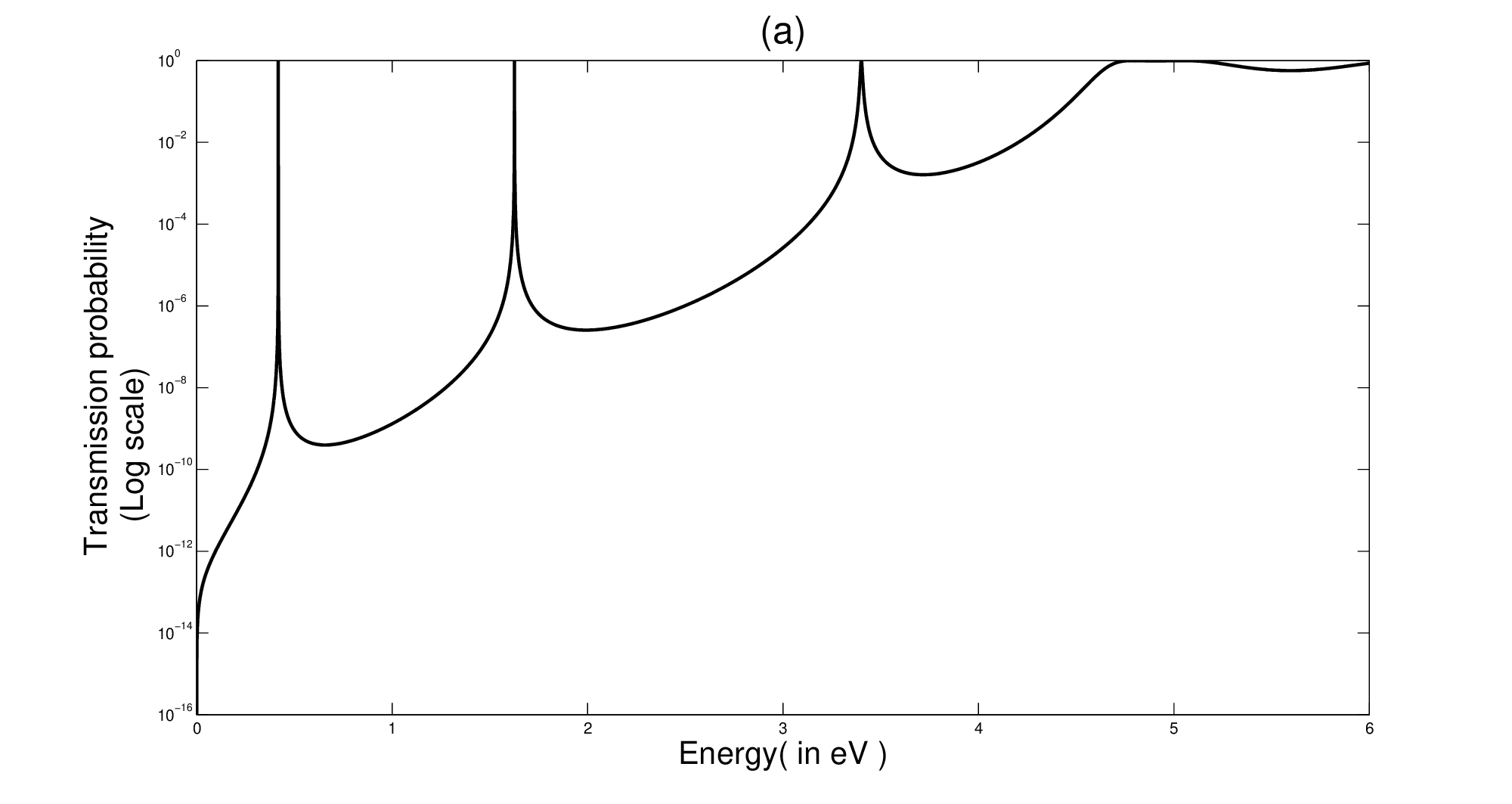}
\includegraphics[width=3.2in,height=2.2in]{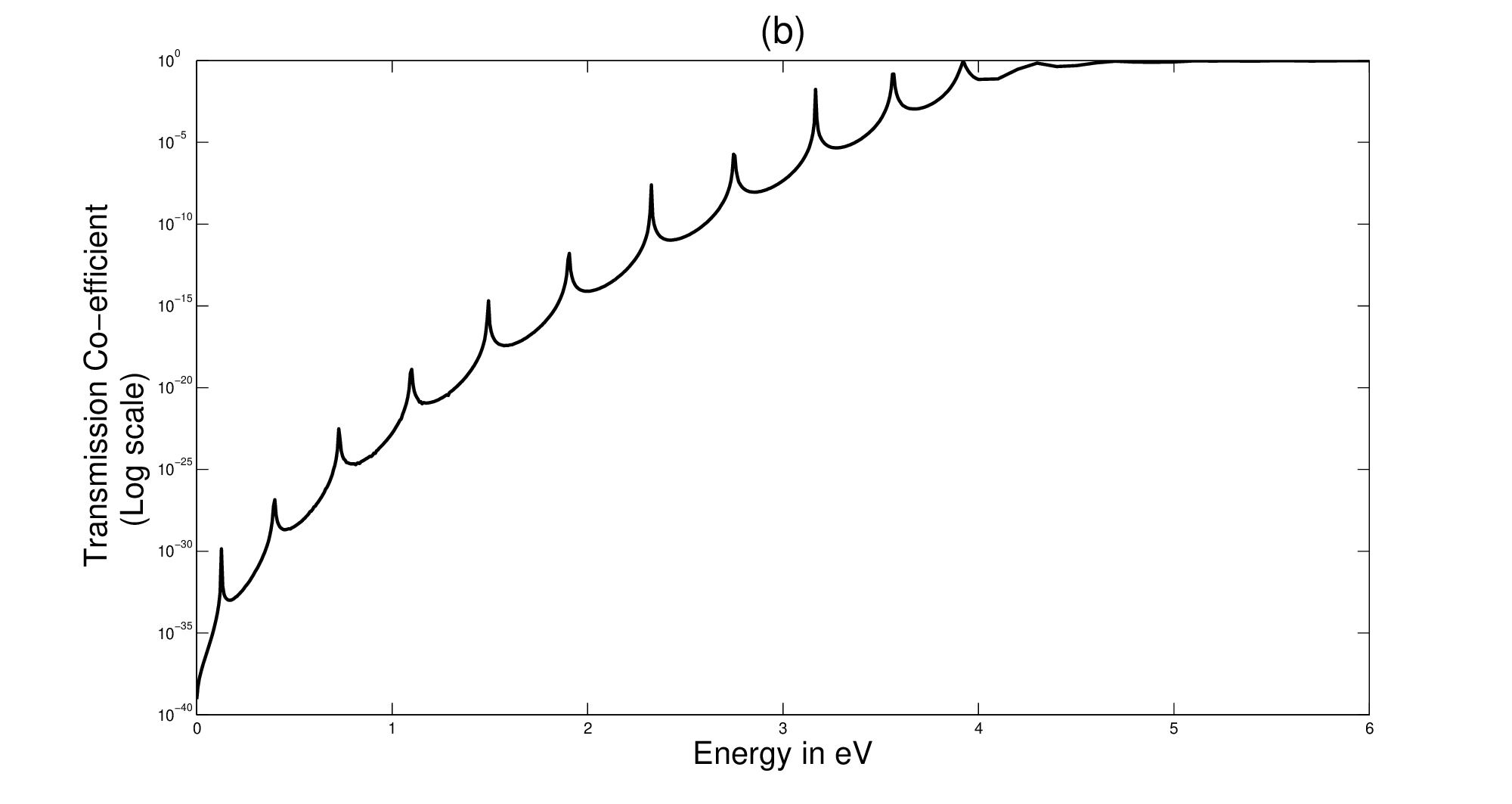}
}
\subfigure{\includegraphics[width=3.2in,height=2.2in]{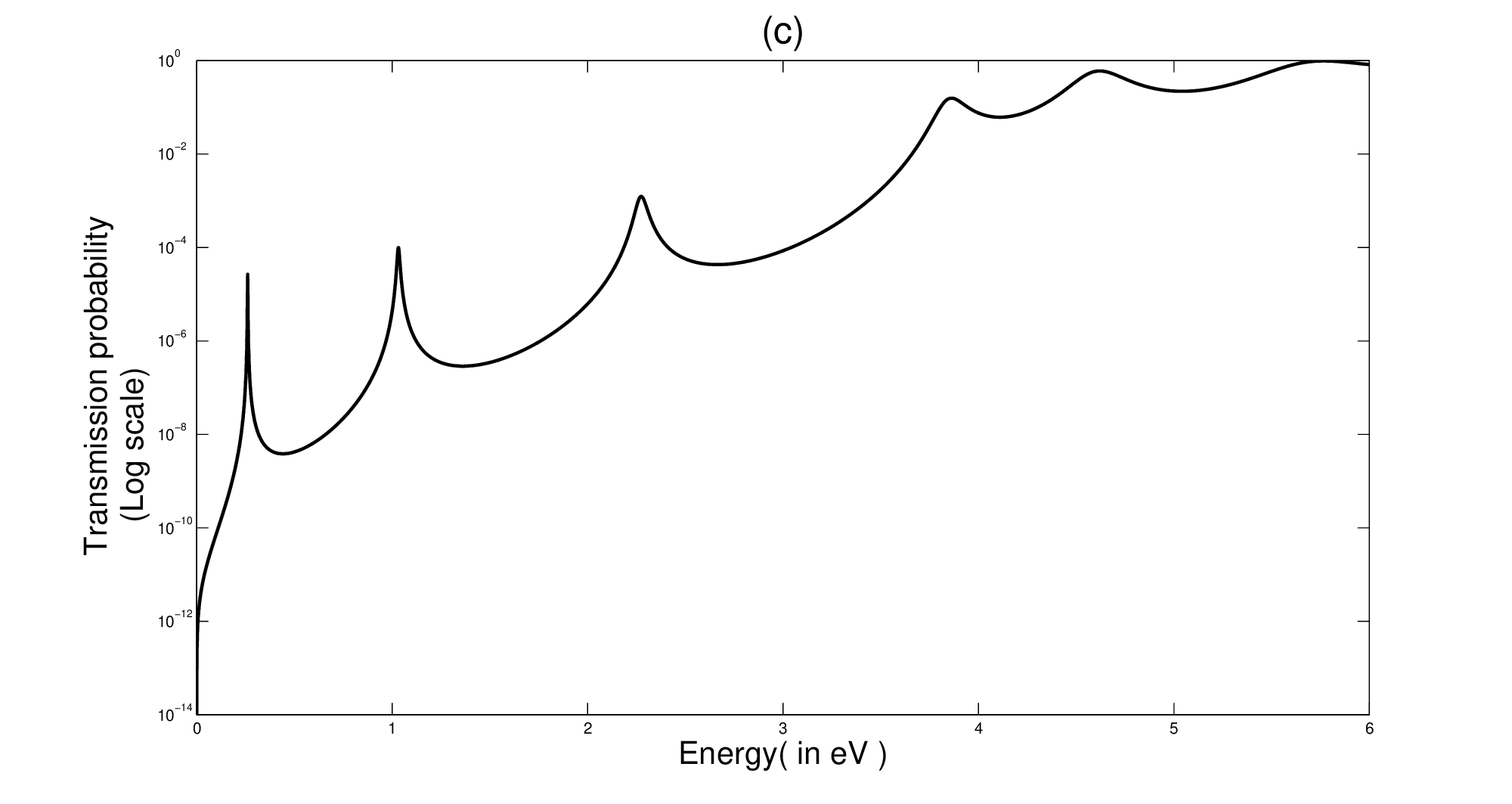}
\includegraphics[width=3.2in,height=2.2in]{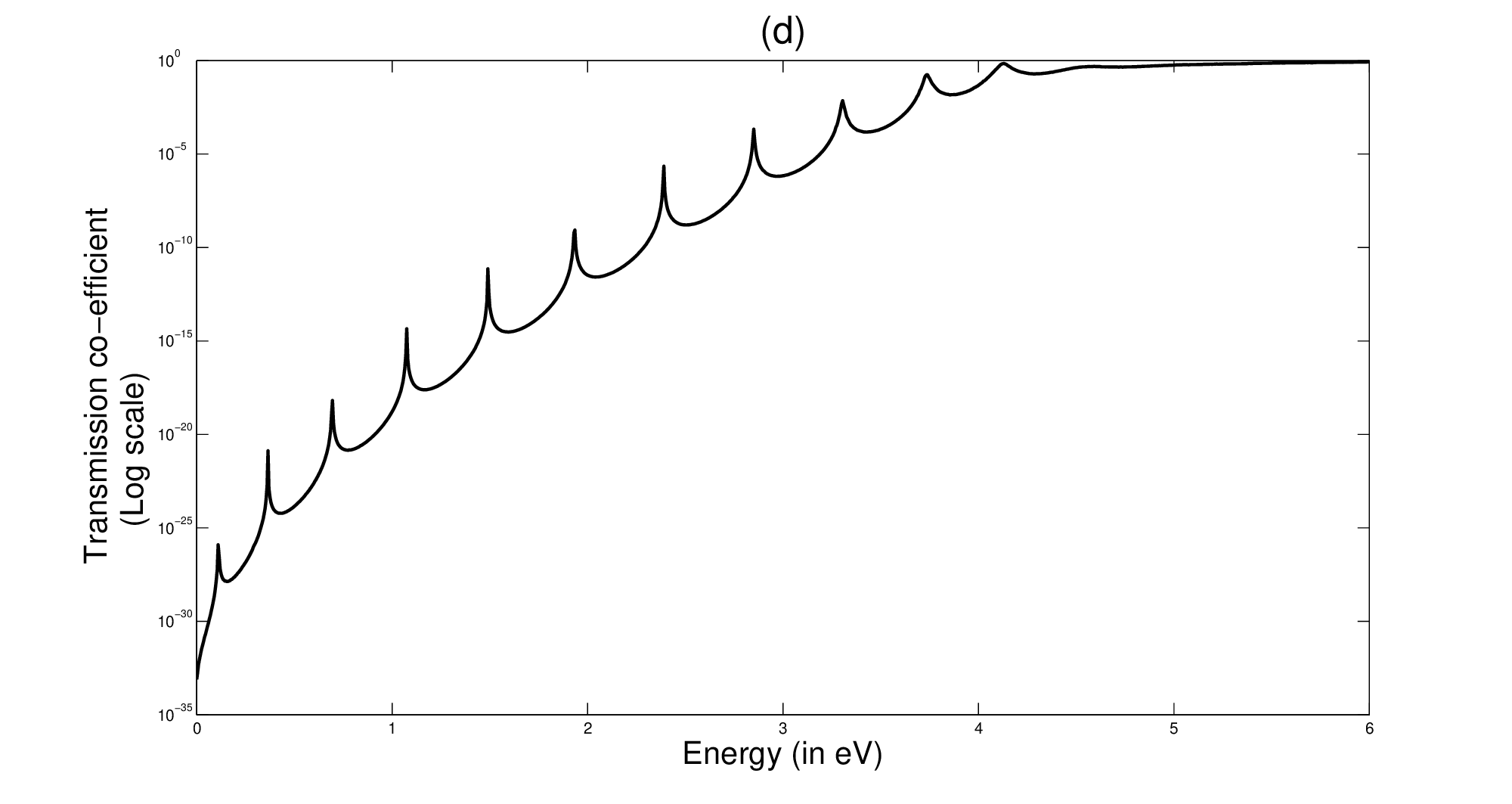}
}
\subfigure{\includegraphics[width=3.2in,height=2.2in]{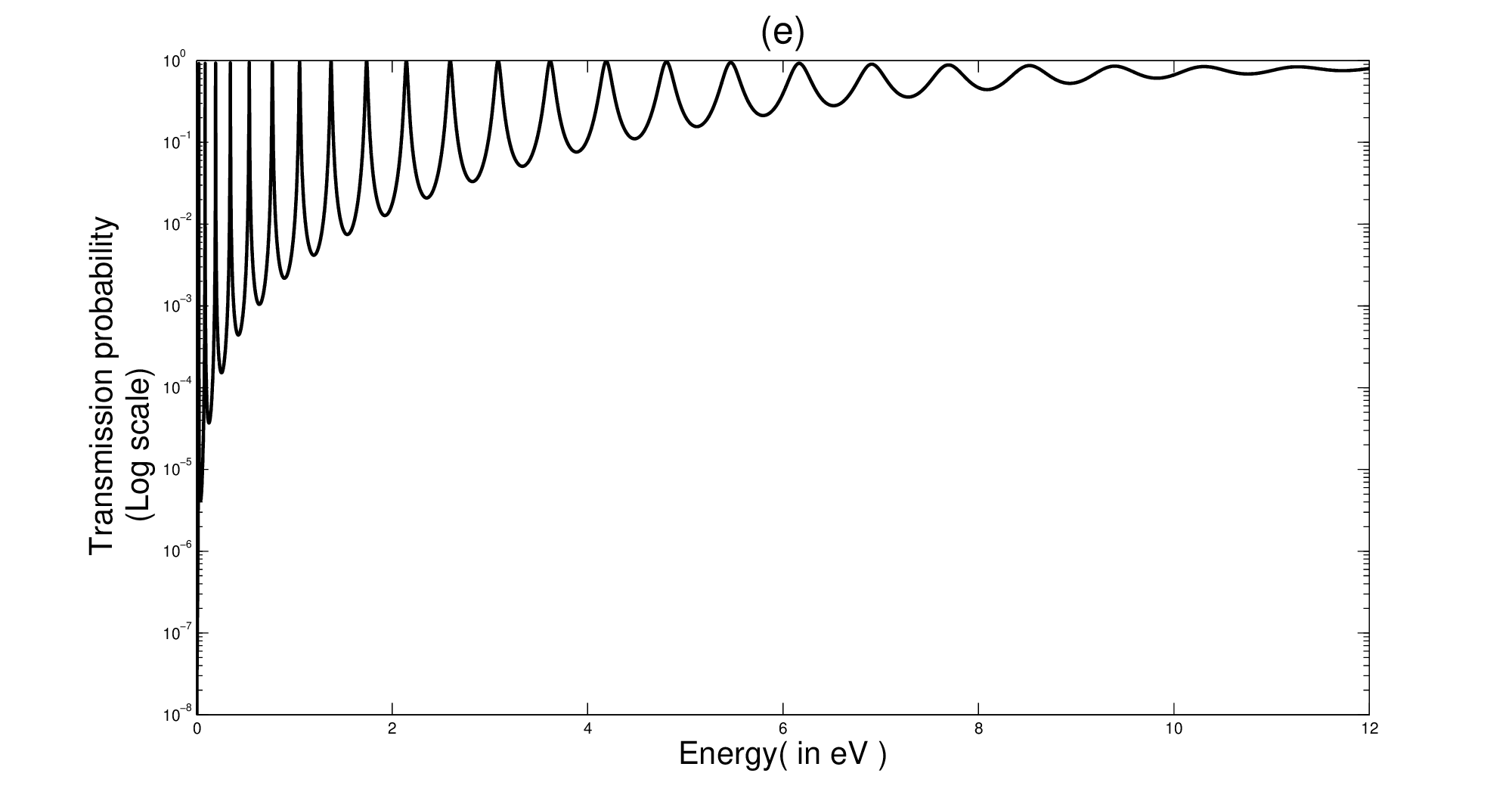}
\includegraphics[width=3.2in,height=2.2in]{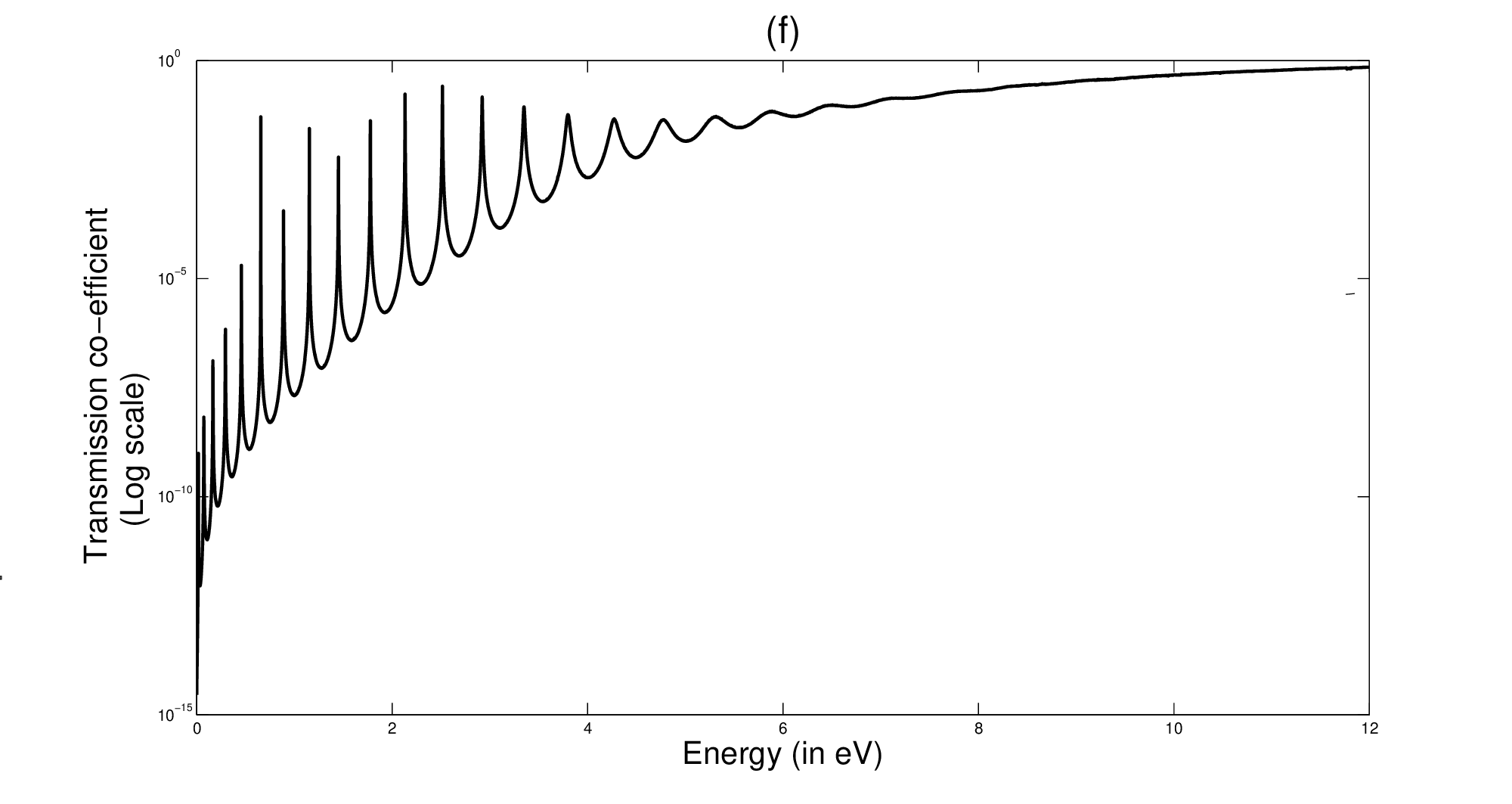}
}
\caption{\footnotesize Comparison of transmission probabilities for the
rectangular and the Gaussian double barrier. 
(a) Square : $V_1 = 4, V_2 = 4$ eV, $w_1 = 0.6, w_2 = 0.6, a = 0.75$ nm
(b) Gaussian : $V_1=4, V_2=4$ \textcolor{black}{eV}, $\sigma_1=0.6, \sigma_2=0.6, a=4.35$ nm
(c) Square : $V_1 = 4, V_2 = 4$ eV, $w_1 = 0.2, w_2 = 0.8, a = 1$ nm
(d) Gaussian : $V_1=4, V_2=4$ eV, $\sigma_1=0.2, \sigma_2=0.8, a=4$ nm
(e) Square : $V_1 = 4, V_2 = 8$ eV, $w_1 = 0.2, w_2 = 0.1, a = 4$ nm
(f) Gaussian : $V_1=4, V_2=8$ eV, $\sigma_1=0.2, \sigma_2=0.1, a=4.9$ nm
The transmission probability asymptotically approaches unity 
for the Gaussian barrier due to the smoothness of the Gaussian barrier potential.
\label{fig:comparison}}
\end{figure}

The $\cal T$ versus $E$ plots in Fig.~\ref{fig:numerical} show the variation of the transmission probability with the energy of an electron incident on a one-dimensional Gaussian double barrier. 
Figure~\ref{fig:numerical} illustrates how the
transmission coefficient changes due to the presence of two barriers, a change in the barrier separation, and a change in the height of one barrier. The energy of the electron is assumed to be of the order of a few electron volts.
In Fig.~\ref{fig:numerical}(a) the two barriers are very close to each other
and the number of resonances is less (Fig.~\ref{fig:numerical}(b)) 
than obtained for larger $a$. 
In Fig.~\ref{fig:numerical}(c) the separation between the barriers is \textcolor{black}{increased, the} well becomes 
wider than the well in 
Fig.~\ref{fig:numerical}(a), and consequently,
the number of resonances increases (Fig.~\ref{fig:numerical}(d)). 
Figure~\ref{fig:numerical}(e) shows two barriers that are so close that they appear to be a single continuous barrier, 
and hence
no resonances (Fig. ~\ref{fig:numerical}(f)) are seen. 
In Fig.~\ref{fig:numerical}(g) we
consider asymmetrical barriers for which $V_{1} \neq V_{2}$ and/or $\sigma_{1} \neq \sigma_{2}$ with $\sigma_{1} V_{1} = \sigma_{2} V_{2}$.
We note that
the effect is predominantly that of a single barrier though resonances 
do appear (Fig. ~\ref{fig:numerical}(h)), signaling the presence of a second 
barrier of much smaller height. 

Next, we compare the transmission coefficients for the
Gaussian and the rectangular double barrier (see Fig.~\ref{fig:comparison}).
We choose $a(\mbox{Gaussian})=3w_1 + 3w_2 + a (\mbox{rect})$, $\sigma_{1}=w_1 $, and $\sigma_{2}=w_2$. We also vary the height of the barriers keeping the (width $\times$ height) for both the barriers the same. Figure~\ref{fig:comparison}(a),(c),(e)  shows the transmission coefficient
for a double rectangular barrier, and Fig.~\ref{fig:comparison}(b),(d),(f) shows the same for a double Gaussian. The qualitative nature of the plots for the rectangular and the Gaussian barriers is similar.
Tunneling occurs for both barriers, though there are differences
that may be used to determine the barrier shape from plots of
$\mathcal T$ versus $E$.

The number of peaks in $\mathcal T$ is large
for both cases. However, the peaks
and valleys are more prominent for the rectangular barrier. The transmission coefficient for the Gaussian does not reach unity (except asymptotically)
and the minima are also at lower values. The cause of this quantitative
difference is the smoothness of the Gaussian barrier.

The following observations are valid if $ a/\sigma$ is large ($\geq 4$). 
The ${\mathcal T}$ versus $E$ plot (Fig.~\ref{fig:comparison})
depends strongly on the product of width and height for any barrier. If we keep the width $\times$ height constant, the graphs are qualitatively similar.

The plots show a large number of valleys and peaks, corresponding to resonant states. Resonant states are a result of the destructive interference between 
waves reflected from the two barriers. The valleys in the $\mathcal T$ versus 
$E$ plots correspond to quasi-bound states.
As the separation-to-width ratio of the barriers is increased, the number of resonances increases, which means that the well can accommodate a larger number of quasi-bound states. If the well width is made indefinitely wide, the resonant states merge into a continuum.

The peaks become sharper as the height of the barriers is increased; i.e. the full width at half maximum of the peaks decreases. If $\Delta$ is the full width at half maximum of the peak in ${\mathcal
T}(E)$, then $\Delta$ decreases as $V_{0}$ and $ V_{1}$ increase.
The peaks also become sharper if the widths of the barriers are increased, which makes it difficult to obtain numerical solutions. The lower the energy at which the peak occurs, the steeper is the peak.
If $E \ll V_{0},\ V_{1}$, $\Delta$ is very small and vice versa.

\subsection{WKB Analysis}
\begin{figure}[h!]
\centering
\subfigure{\includegraphics[width=3.2in,height=3in]{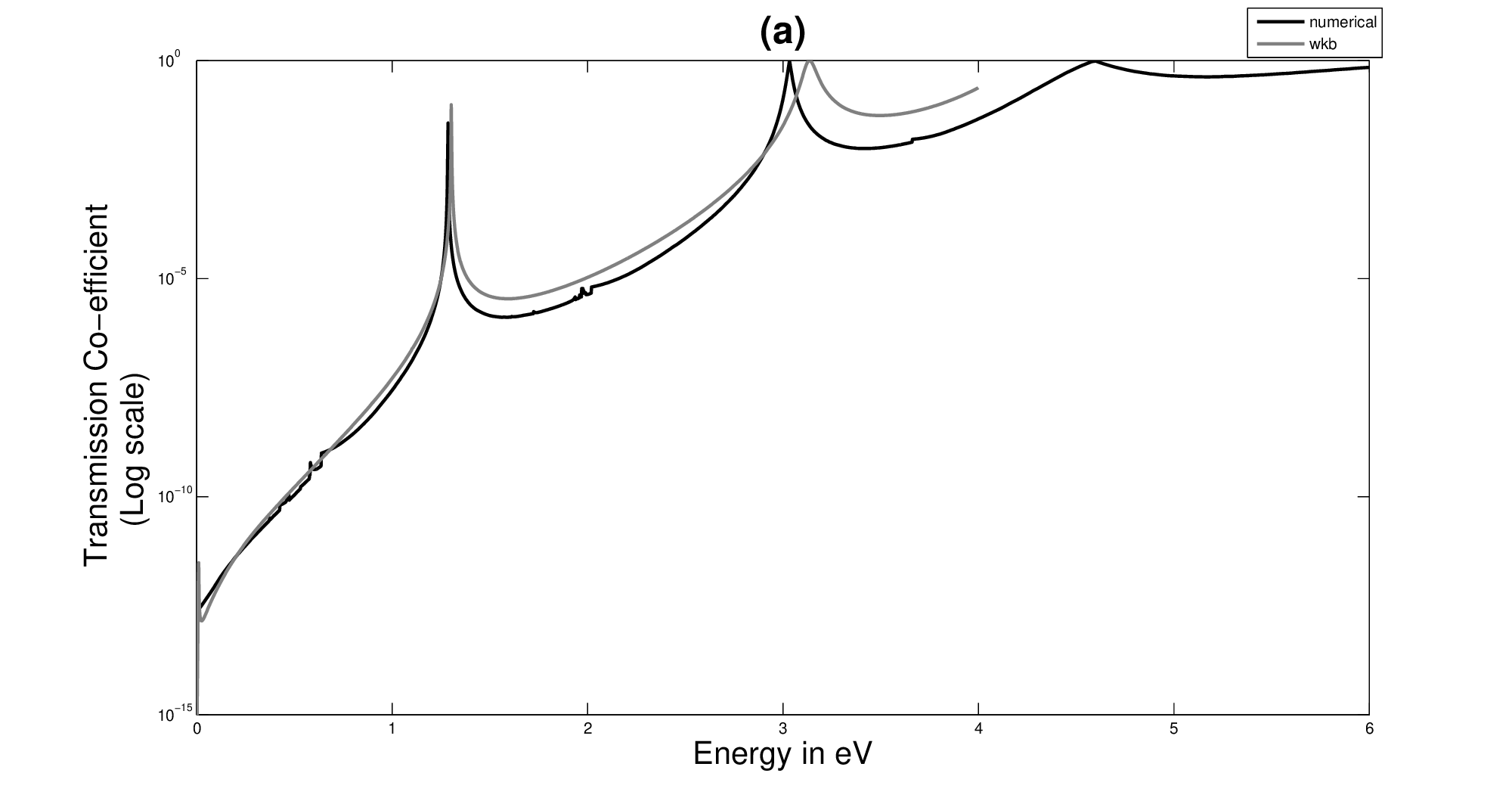}
\includegraphics[width=3.2in,height=3in]{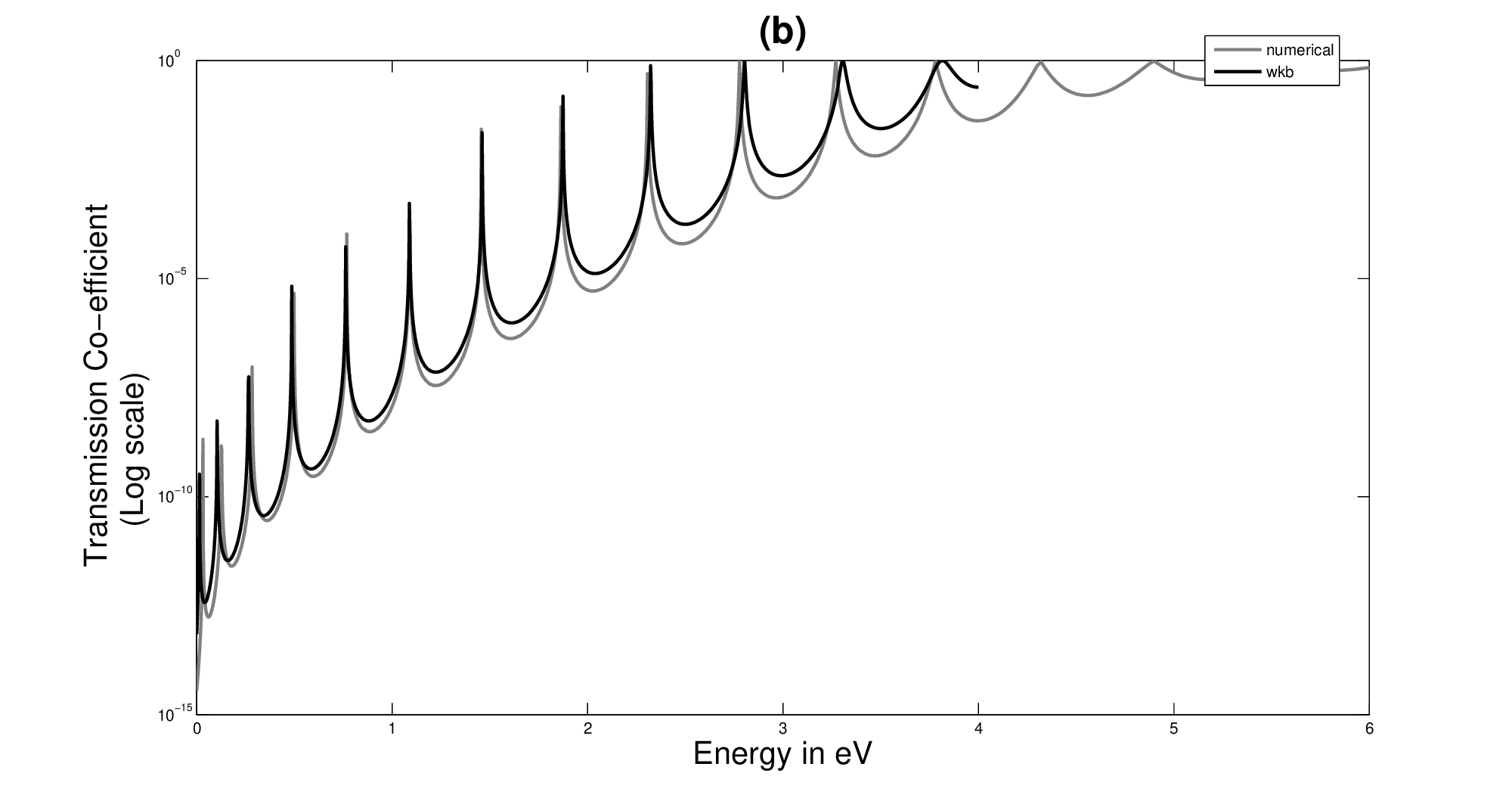}}
\subfigure{\includegraphics[width=3.2in,height=3in]{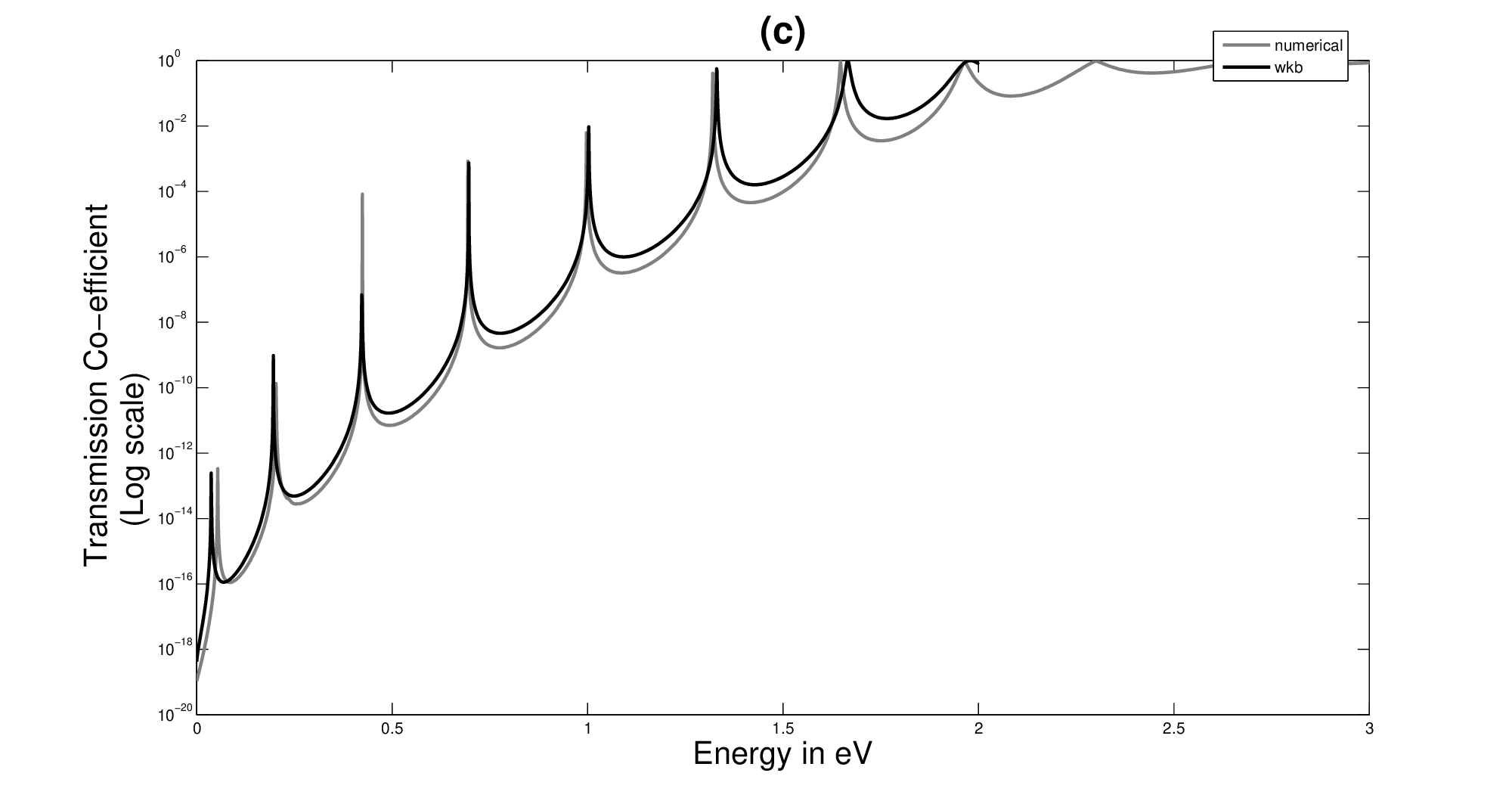}
\includegraphics[width=3.2in,height=3in]{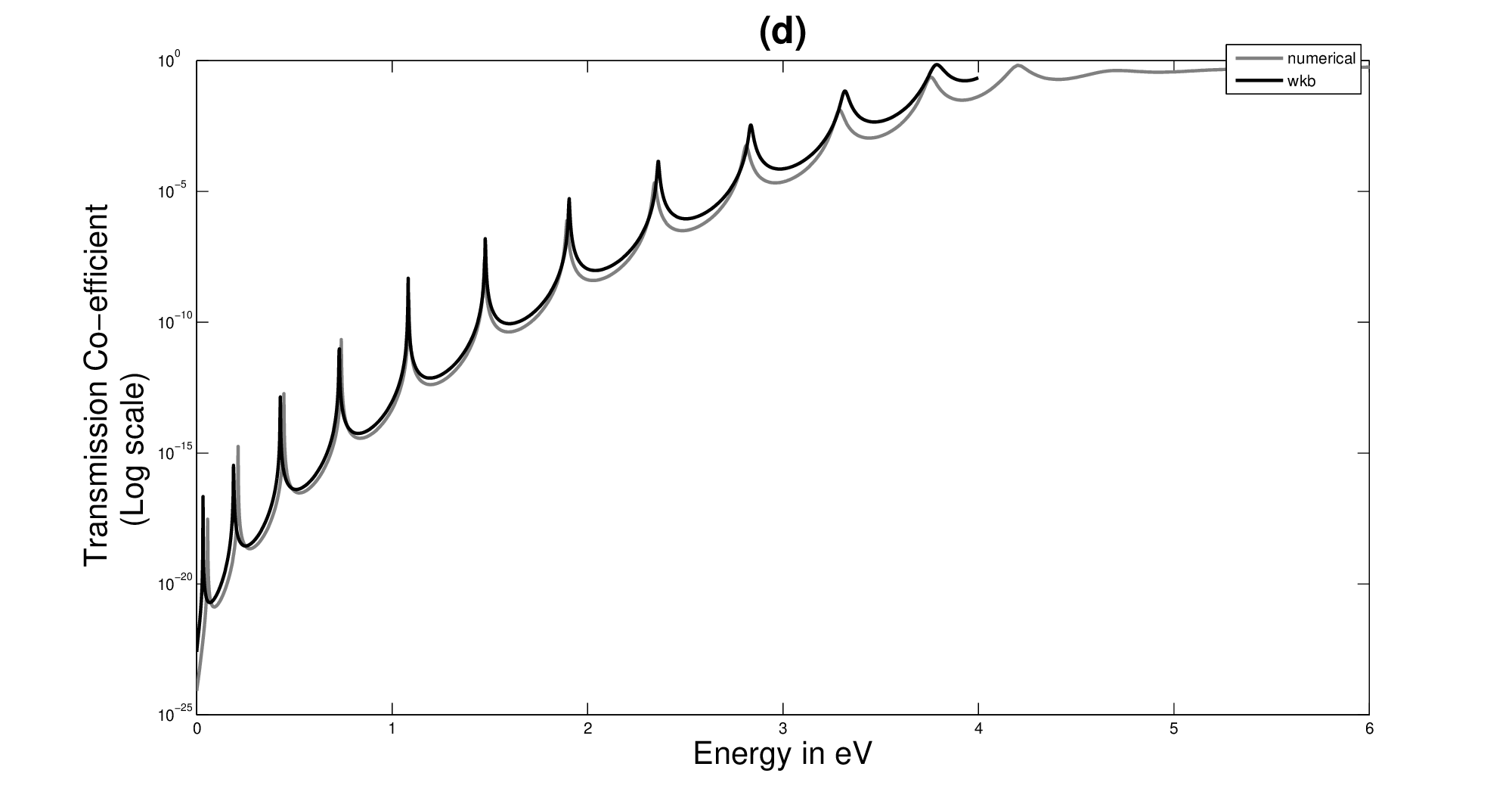}}
\caption{Comparison of the transmission probabilities calculated by numerical solution of Schr\"{o}dinger equation (gray) 
and the WKB approximation (black) for a Gaussian double barrier with parameters: 
(a) $V_1=V_2=4$\,eV, $\sigma_1=\sigma_2=0.2$\,nm, $a=1$\,nm,
(b) $V_1=V_2=4$\,eV, $\sigma_1=\sigma_2=0.2$\,nm, $a$ is increased to $4$\,nm.
(c) $V_1=V_2=2$\,eV, $\sigma_1=\sigma_2=0.4$\,nm, $a=4$\,nm. The potential barrier height is larger for this barrier.
(d) $V_1=V_2=4$\,eV, $\sigma_1=0.6,\, \sigma_2=0.1$\,nm, $a=4$\,nm. In this case we consider an asymmetrical barrier.
The results obtained by the two methods are in good agreement.}
\label{fig:comparison1}
\end{figure}

The WKB method provides a way
of obtaining analytical expressions for the transmission coefficient. In Figure~\ref{fig:comparison1}
we compare the results obtained for the transmission probability using the semiclassical analysis method and the numerical solutions\textcolor{black}{.}
Figures~\ref{fig:comparison1}(a) and (b) 
differ because of the larger value of $a$ in Fig.~\ref{fig:comparison1}(b)
(all other parameters are the same). 
In Fig.~\ref{fig:comparison1}(c) the heights are halved, and the
widths are doubled compared to Fig.~\ref{fig:comparison1}(b). In
Fig.~\ref{fig:comparison1}(d) the Gaussian barrier is asymmetrical with $\sigma_{1}/\sigma_{2} = 6$.

For $E \ll V_{1}$ the values of $\mathcal{T}$ are almost the same, or at 
least of the same order of magnitude using either the numerical method or 
the WKB approximation. The greatest deviations from the numerical solution 
are observed for energies near $V_{0}$. This deviation is expected because 
the WKB approximation is not valid near the classical turning points. The 
difference between the WKB method and the numerical solution of Schrodinger's 
equation also becomes prominent when the separation between the barriers 
becomes comparable to the width of the barriers. In the calculation of the 
turning points we neglected the contribution of the first barrier to the 
turning points of the second barrier and vice versa so that we could obtain 
analytic expressions for the classical turning points;
otherwise, we would have to solve the transcendental equation $V(x)= [V_1 \exp(-x^2/ 2\sigma_1^2) + V_2 \exp(-(x-a)^2/ 2\sigma_2^2)] = E$ for each value of 
$E$. 
Thus the difference between the semiclassical results and the numerical solutions for the transmission probability 
becomes prominent when the barriers are close to each other, but the assumption yields acceptable results in return for a shorter 
computation time.

The WKB approximation becomes more accurate as the separation between the 
barriers increases. 
The numerical values of ${\mathcal T}(E)$ versus $E$ in Fig.~\ref{fig:comparison1} do not match the WKB approximation 
results for energies $ E > \min (V_{1}, V_{2})$. 
For asymmetrical barriers with non-identical potential heights, 
the WKB approximation breaks down for $E > \min(V_{1}, V_{2})$ because the 
formulation we have used involves four classical turning points. The potential 
function 
reduces to a single barrier with two turning points for $E>\min(V_1, V_2)$. 
However, the WKB results agree quite well for $E < \min(V_{1},V_{2})$.

The advantages of the WKB method are that it requires about an order of magnitude less computation time than the direct numerical 
solution of Schrodinger's equation. The main limitation of the WKB method is that it relies heavily on the turning points, 
and hence is difficult to apply to energies greater than the potential height of any of the barriers. 

\section{Tunneling Time}

With the advent of the fabrication of nanometer semi-conductor structures, the calculation of
tunneling time has acquired
more importance.\cite{davies} For one-dimensional single barriers, the 
tunneling time becomes independent of the barrier width for opaque barriers. 
This ``Hartman effect''\cite{hartman} predicts superluminal and arbitrarily 
large group velocities inside sufficiently long barriers. There is some 
controversy concerning the definition of the tunneling time.\cite{haugestoveng} The expression for the tunneling time we have used is variously named the ``group delay'' or ``phase time.''

If we assume a single rectangular barrier, with the potential
\begin{equation}
V(x)=\begin{cases}
 V_{0} & (0 < x < L) \\
 0 & \mbox{elsewhere}, \\
\end{cases}
\end{equation}
the wavefunction may be written as\cite{haugestoveng}
\begin{equation}
\psi(x)=\begin{cases}
\exp(ikx) + \sqrt{\mathcal{R}}\exp(-ikx+i\gamma) & (x<0) \\
\psi_1(x) & (0 \leq x \leq L) \\
\sqrt{\mathcal{T}}\exp(ikx+i\alpha) & (x>L). \\
\end{cases}
\end{equation}
where $\mathcal{T}$ and $\mathcal{R}$ are the transmission and reflection probabilities, $k$ the wave vector, and $\alpha$ and $\gamma$ are phase constants dependent on $k$.
If we apply the stationary phase approximation to the peak of the transmitted wave packet, $\partial \{\mbox{arg}(\psi(x,t))\} /\partial k = 0$, we find 
\begin{equation}
\frac{d\alpha}{dk} -\frac{1\, dE}{\hbar\ dk}t = 0.
\end{equation}
Thus, the temporal delay may be defined as
\begin{equation}
\label{grpdelay}
\tau = \hbar\frac{\partial\alpha}{\partial E}.
\end{equation}
From the transmission amplitude for a single barrier we find the phase shift to be $\alpha = -\tan^{-1}((\kappa^{2}-k^{2}) \tanh(\kappa L) /2\kappa k)$,
where $\kappa=\sqrt{2m(V_{0}-E)}/\hbar$ and $k=\sqrt{2mE}/\hbar$. By using Eq.~(\ref{grpdelay}) we have
\begin{equation}
\tau_{g} \stackrel{\kappa L\rightarrow \infty}{\longrightarrow} \frac{2m}{\hbar k \kappa},
\end{equation}
where the subscript $g$ denotes group delay.

For a rectangular double barrier the transmission amplitude is expressed by Eq.~(\ref{eq:sqtrans}), and the phase time is\cite{olkhrecami,petrilloolkh}
\begin{equation}
\label{eq:eqn20}
\tau_{g} = \hbar \frac{\partial}{\partial E}\arg [T \exp(ik_{1}b)].
\end{equation}
In the limit of $k_{2}w_1$ and $ k_{3}w_2 \to \infty$, the phase time becomes independent of the barrier width and separation:
\begin{equation}
\label{eq:taug1}
\tau_{g} = \frac{2m}{\hbar k_{1} k_{2}},
\end{equation}
except at resonance and anti-resonance.

To illustrate what happens at resonance and anti-resonance,\cite{hgwinful} we take $E = V_{1}/2$ and $V_0 = V_{1}=V_{2}$. When the phase shift $2k_{1} a = 2m\pi$ ($m=1,2,3,\ldots$) (resonance), the group delay becomes
\begin{equation}
\label{eq:taug2}
\tau_{g}^{\rm res} = \frac{(1+R_{0})}{T_{0}} \frac{a}{v},
\end{equation}
where $v = p/m = \sqrt{2mE}/m$, and $ R_{0}$ and $T_{0}$ denote the reflection and transmission probabilities through a single barrier of width $w_1$. 
At anti-resonance ($2k_{1} a =(2m+1)\pi$),
\begin{equation}
\label{eq:taug3}
\tau_{g}^{\mbox{\tiny anti-res}} = \frac{T_{0}}{(1+R_{0})} \frac{a}{v},
\end{equation}
indicating that the phase time increases linearly with the inter-barrier separation, $a$, at both resonance and anti-resonance.

\subsection{Tunneling times for general double barrier potentials}
The wave function for $x>x_4$ (Fig.~\ref{fig:wkb}) 
is given in the WKB approximation by
\begin{equation}
\psi_{V} = A |p|^{-1/2} \exp\left(\frac{i}{\hbar} \int_{x_{4}}^{x}|p|\,dx + i\frac{\pi}{4}\right),
\end{equation}
and the incident wavefunction is 
\begin{equation}
\psi_{\rm incident} = A|p|^{-1/2}\left\{i(C_{3}+C_{4})/T_{3} + iT_{3}(C_{4}-C_{3})/4 \right\}\exp\left(-i\!\int_{x}^{x_{1}} p/\hbar \, dx - i\pi/4 \right).
\end{equation}

In analogy with Eq.~(\ref{eq:eqn20}), the group delay is found to be
\begin{equation}
\label{eq:taug}
\tau_{g} = \hbar \frac{\partial}{\partial E}\arg\left[(\psi_{V}/\psi_{\rm incident}) \exp\left(\int_{x_{1}}^{x_{4}}|p|/\hbar \, dx\right) \right].
\end{equation}
If we substitute $\psi_{V}$
and $\psi_{\rm incident}$, into Eq.~\ref{eq:taug} 
we obtain
\begin{equation}
\tau_{g} = \hbar \frac{\partial}{\partial E}\arg\left [\frac{T_{3}(C_{4}-C_{3})}{4} + \frac{C_{4}+C_{3}}{T_{3}}\right ]^{-1},
\end{equation}
and after some algebra
\begin{subequations}
\label{eq:tauwkb}
\begin{align}
\tau_{g} & = \hbar \frac{\partial}{\partial E} \tan^{-1}\frac{2 \tan T_{2}}{T_{1}^{2}} \\
& = \frac{2}{T_{1}^{4} + 4 \tan^{2} T_{2}} T_{1}^{2}\left(\sec^{2} T_{2} \frac{\partial T_{2}}{\partial E} - 2\frac{\tan T_{2}}{T_{1}} \frac{\partial T_{1}}{\partial E}\right).
\end{align}
\end{subequations}

Equations ~\ref{eq:taug1}--\ref{eq:taug3}, which are applicable for a rectangular barrier, can be recovered from Eq.~(\ref{eq:tauwkb}) as follows. We have $T_{1}=\exp(-k_{2} w_1)$ and $T_{2} = k_{1} a$. We let $\phi= \tan^{-1}(2\tan (T_2)/T_{1}^{2})$ and write
$\tau_{g} = \hbar \partial \phi/\partial E$. We find that the phase tunneling time through a rectangular double barrier of width $w_1$ and separation $a$ is
\begin{equation}
\tau_{g} =2\cos^{2} \phi \frac{m}{\hbar}\exp(2k_{2}w_1)(a\sec^{2} (k_{1} a)/k_{1} - 2w_1\tan (k_{1} a)/k_{2}).
\end{equation}
To illustrate the behavior of $\tau_g$ near a resonance, we assume $E=V_{1}/2$ so that $k_{1}=k_{2}$. At resonance, $2k_{1} a=2m\pi$, where $m$ is an integer. The group delay at resonance assumes the form
\begin{equation}
\tau_{g}^{\rm res} = 2e^{2k_{2} w_1} a /v \approx (1+R_{0})a/T_{0}v,
\end{equation}
because $\cos\phi = 1$ and $\tan k_{1} a =0$. Here, $R_{0}=(1/T_{1}-T_{1}/4)^{2}/(1/T_{1}+T_{1}/4)^{2}$ and $T_{0}=(1/T_{1}+T_{1}/4)^{-2}$ are the reflection and transmission probabilities through a single barrier ($(1+R_{0})/T_{0} = 2(T_{1}^{-2} + T_{1}^{2}/16) \approx 2e^{k_{2} w_1}$).

\begin{figure}[h!]
\centering
\subfigure{\includegraphics[width=3.4in,height=2.8in]{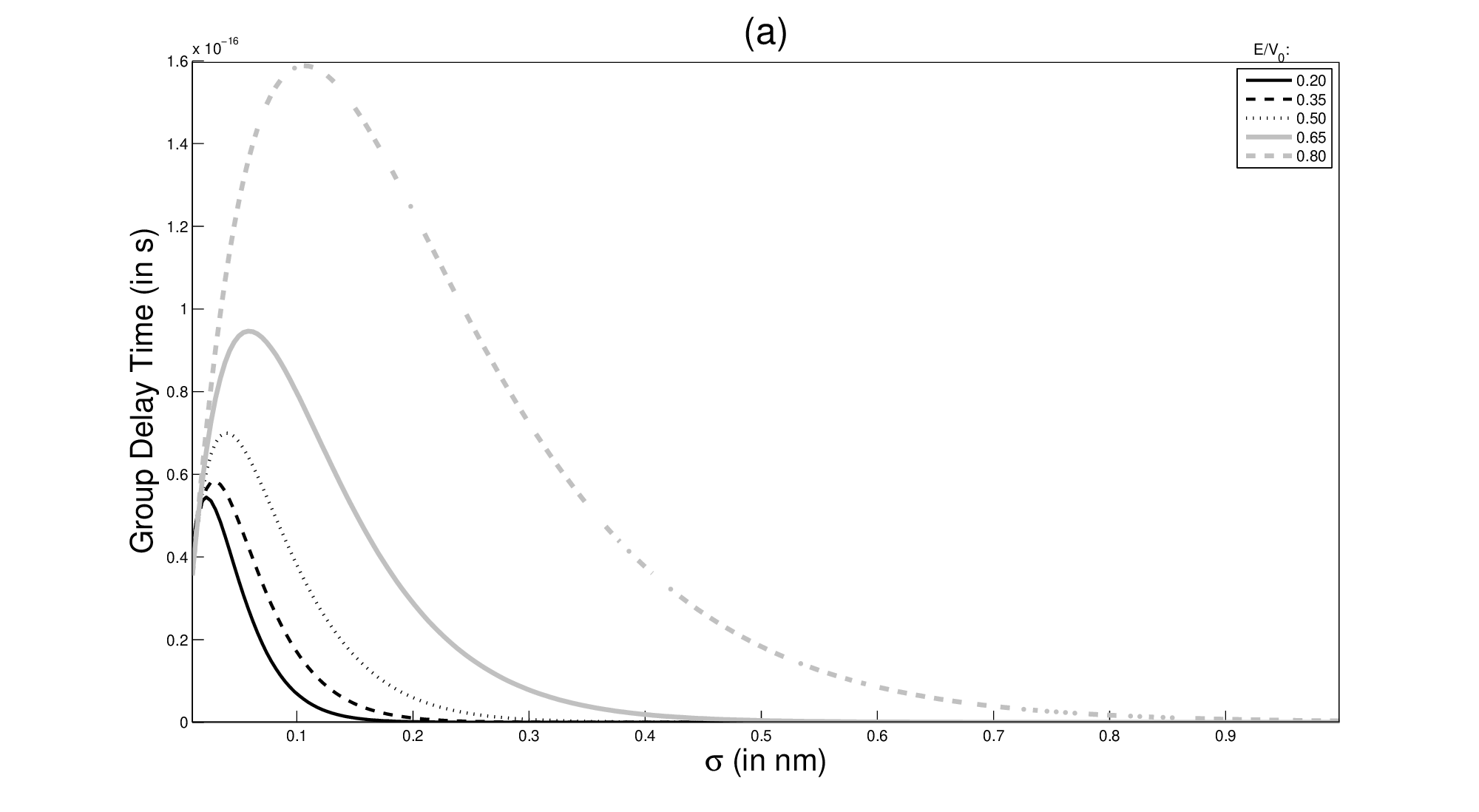}
\includegraphics[width=3.4in,height=2.8in]{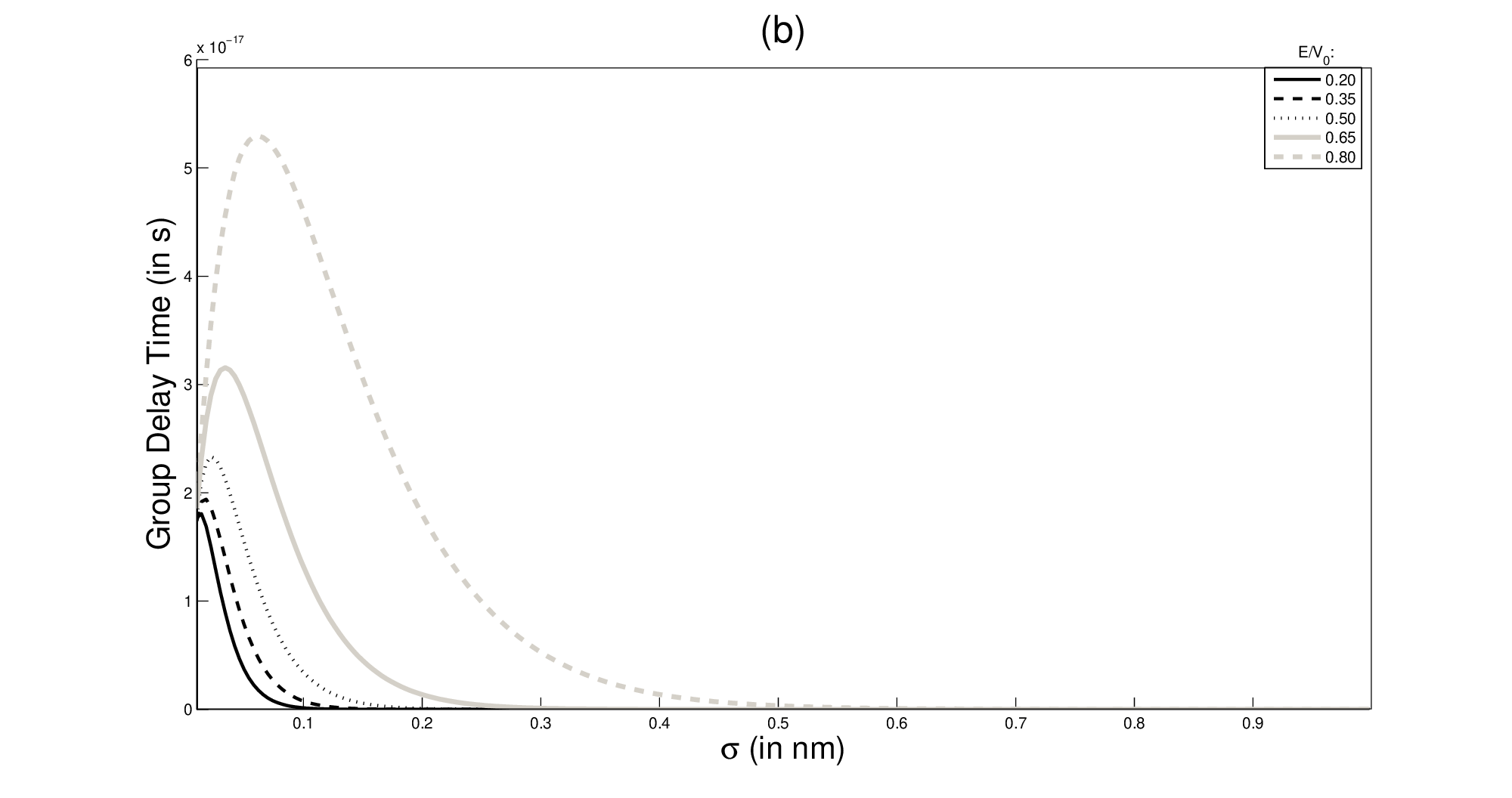}}
\caption{Variation of tunneling time with barrier width $\sigma$ for a double Gaussian barrier, with
(a) $V_0=4$\,eV; $a=8$\,nm
(b) $V_0=12$\,eV; $a=8$\,nm, for various energies. 
The tunneling time increases up to a certain value of $\sigma$ and then decreases rapidly. The maximum in the curve
shifts to larger values of the barrier width for higher energies.} 
\label{fig:tauvssigma}
\end{figure}

\begin{figure}[h!]
\centering
\subfigure{\includegraphics[width=3.4in,height=2.8in]{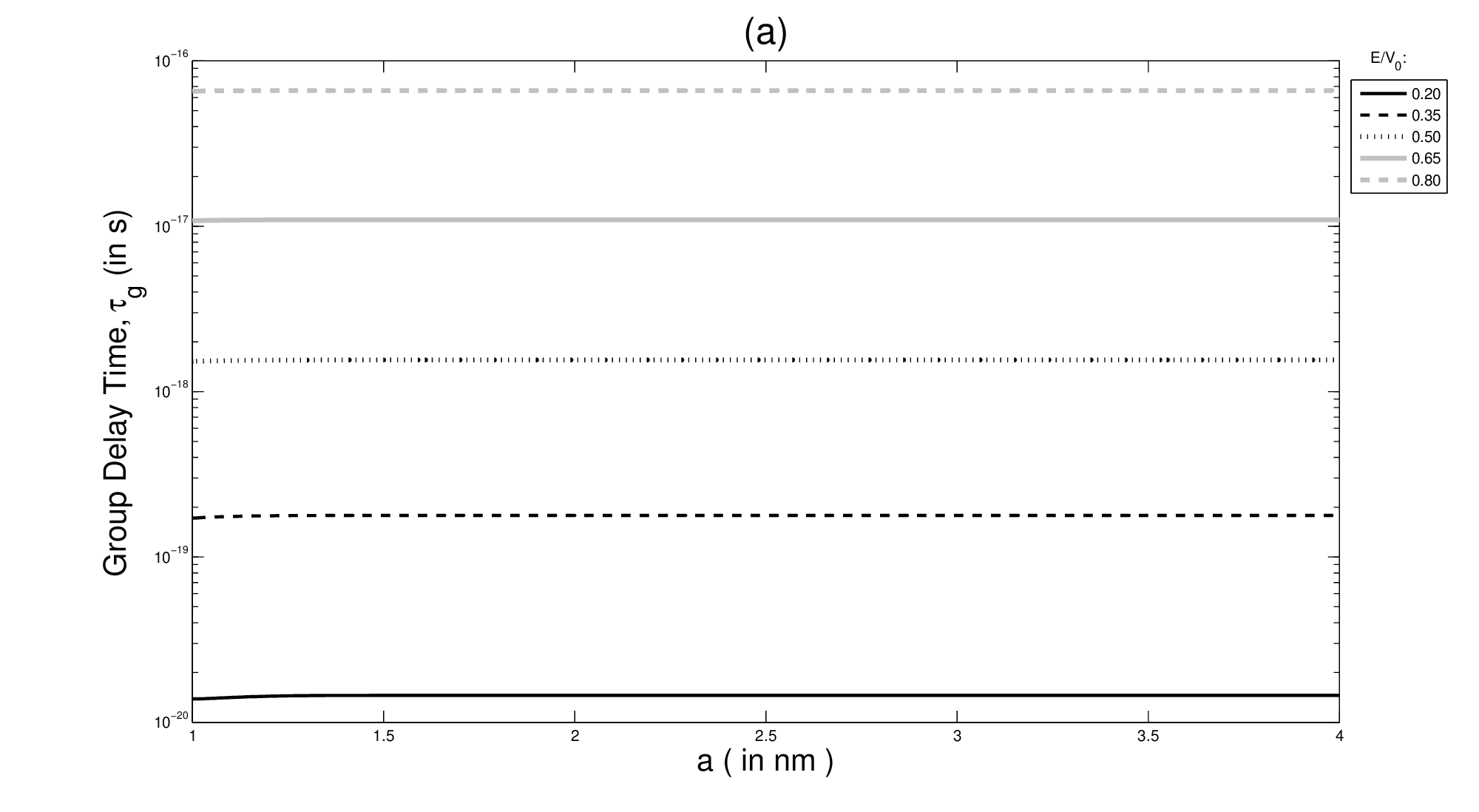}
\includegraphics[width=3.4in,height=2.8in]{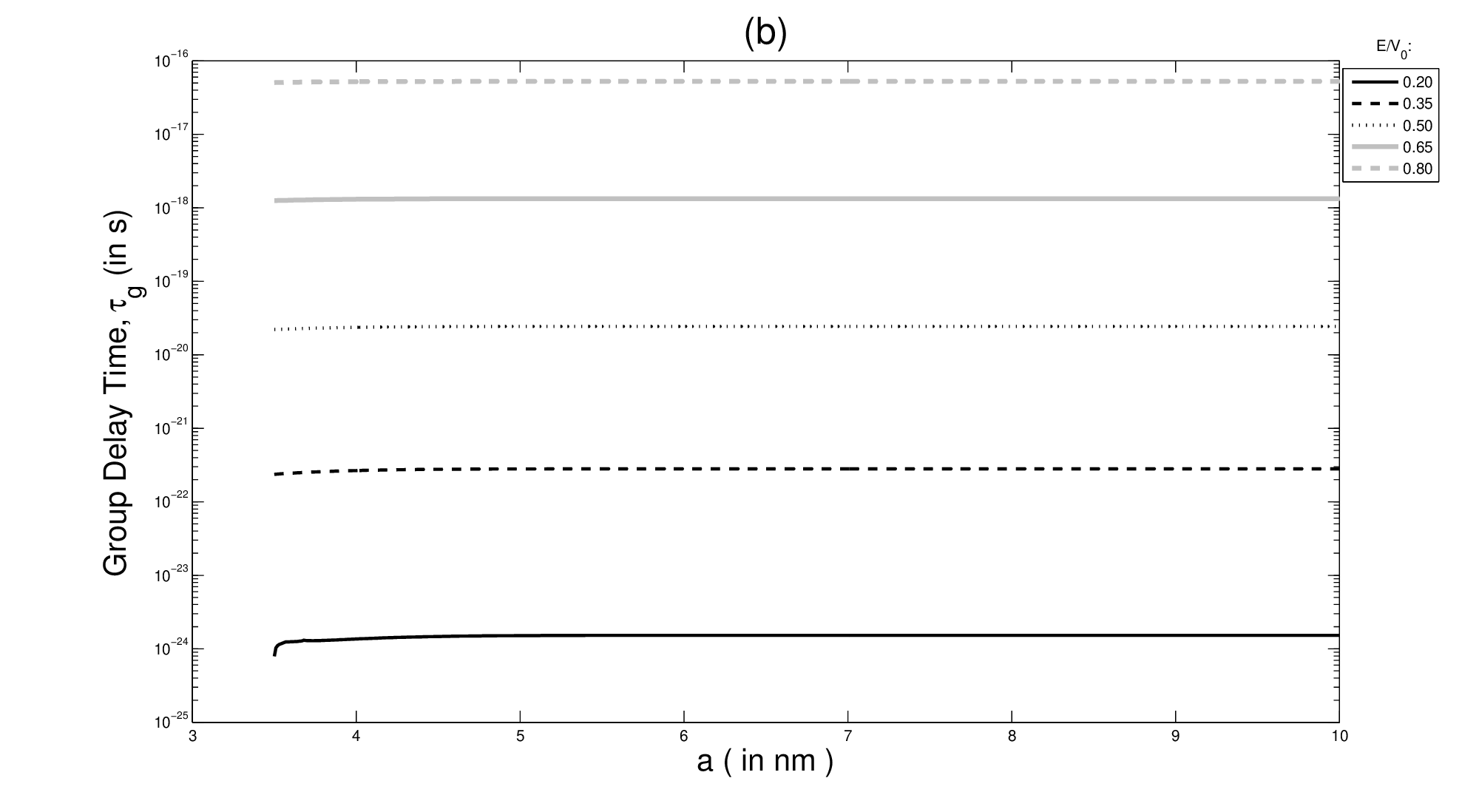}}
\caption{Variation of the tunneling time through a double Gaussian barrier with
inter-barrier distance $a$ for (a) $V_0= 6$\,eV, $\sigma_1=0.2$\,nm, and (b) $V_0=1.50$\,eV$,\sigma_1=0.8$\,nm. Note that the tunneling time varies over several orders of magnitude for various energies of the incident particle.}
\label{fig:tauvsx0}
\end{figure}

Similarly at anti-resonance ($2k_{1} a=(2m+1)\pi)$, $m$ an integer, we have
\begin{equation}
\tau_{g}^{\mbox{\tiny anti-res}} = T_{0} a/v(1+R_{0}).
\end{equation}

We next obtain $\tau_g$ for a Gaussian double barrier.
The relevant integrations for obtaining $T_3$ are performed numerically.
Figure~\ref{fig:tauvssigma} shows the variation of $\tau_g$ with $\sigma$ for fixed $a$. 
Figures~\ref{fig:tauvsx0}(a) and (b) show how $\tau_g$ varies with $a$ for fixed $\sigma$.

We see that the group delay increases with an increase in the barrier width up to a certain value and then falls off rapidly on widening the barrier further. The peak in the group delay versus barrier width shifts toward the greater widths at higher energies. The maxima in $\tau_{g}$ shift to greater values of $\sigma$ at higher energies. The group delay decreases with increasing $V_{0}$, and for fixed $V_{0}$ $\tau_g$ increases with $E$ if other parameters remain constant. 
The tunneling time is more or less independent of the interbarrier distance as is evident from Fig.~\ref{fig:tauvsx0},
which is consistent with the Hartman effect.

\section{DOUBLE BARRIERS IN HETEROSTRUCTURES}
Double barriers of either type (smooth or square) appear in various physical situations (as
outlined in Sec.~I). We now discuss in some detail, how they
may arise in the context of one such scenario-semiconductor heterostructures.

\begin{figure}[h!]
\centering
\caption{Semiconductor heterostructures and superlattices: (i) Energy band diagram of
an n-N semiconductor heterojunction;
(ii) Heterojunctions can be of three types classified according to the lineup of the conduction
and valence bands: straddled ($E_{C2}>E_{C1}, E_{V2}<E_{V_1}$), staggered ($E_{C2}>E_{C1}, E_{C1}>E_{V2}>E_{V1}$), and 
broken ($E_{C2}, E_{V2}>E_{C1}$);
(iii) An unbiased GaAs/AlGaAs superlattice and the corresponding potential energy diagram;
(iv) A biased superlattice created by the application of an electric field, which creates a gradient in the potential function.
}
\label{fig:heterostructure}
\includegraphics[width=7in]{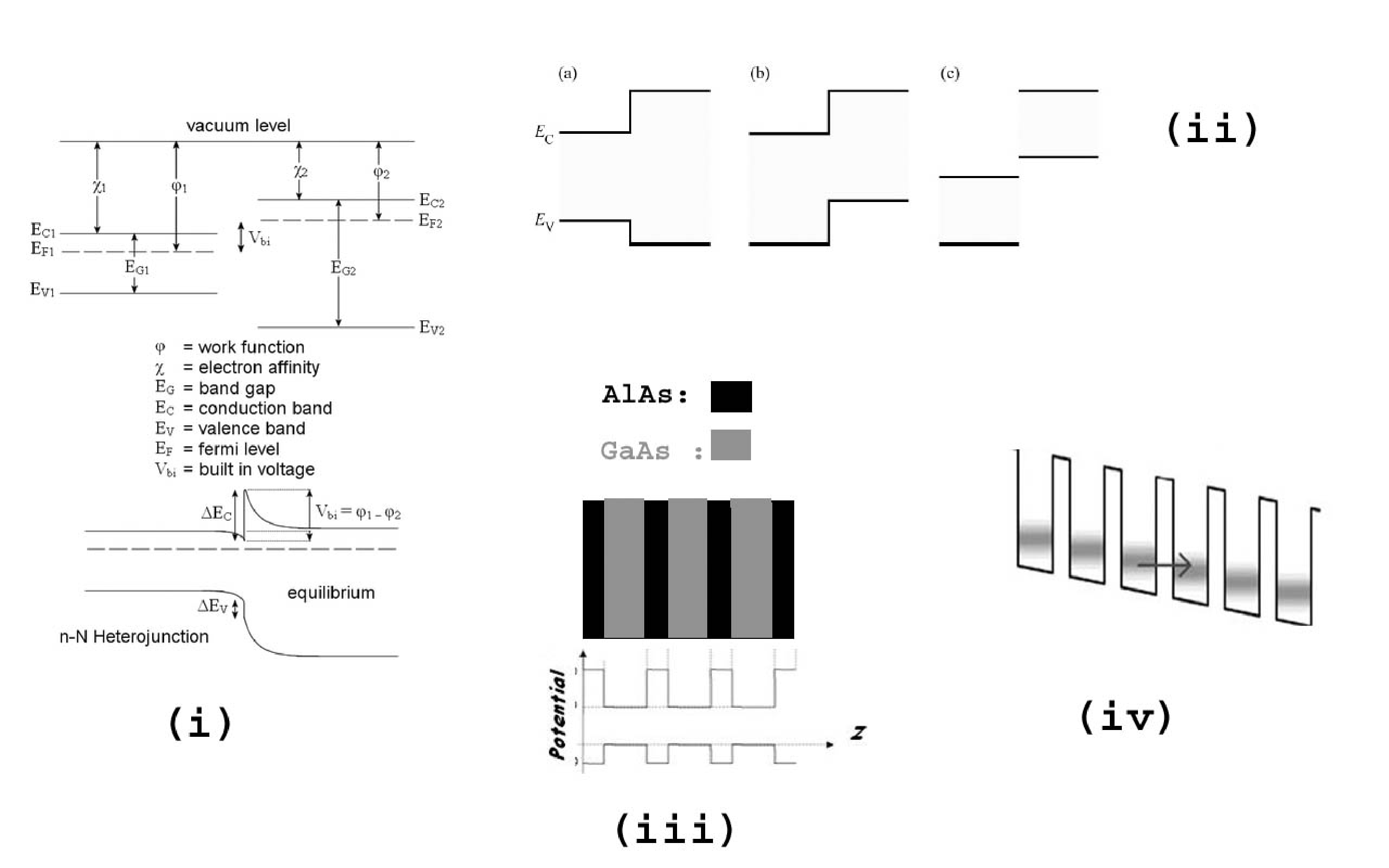}
\end{figure}

A heterojunction (Fig.~\ref{fig:heterostructure}(i),(ii)) is the interface between two dissimilar solid state materials, including crystalline and amorphous structures of metals, insulators, and semiconductors. A combination of one or more heterojunctions in a device is a heterostructure. As discussed in Sec.~I, heterostructures having semiconductor substrates are widely used in the fabrication of devices that have desired electron characteristics as functions of applied potentials.\cite{heterostructures}

Quantum structures may be classified as wells, wires, or dots, depending on whether the carriers are confined in one, two or three dimensions. Semiconductors can be used in all of these structures, but we consider one-dimensional potential wells and barriers, which can be fabricated by growing a single layer of one material (such as GaAs) between two layers of different materials (such as AlGaAs), as in Fig.~\ref{fig:heterostructure}(iii). Other pairs of semiconductors having profound importance in technological applications include the III-V compounds GaInAs/InP, GaInAs/AlInAs, GaSb/InSb and the II-VI compounds CdZnSe/ZnSe, ZnSTeSe/ZnSSe.

In a semiconductor the difference between the conduction and valence band energies of the two materials at the
heterojunction with respect to the vacuum level and the Fermi level is responsible for the formation of quantum wells,
whose heights are of the order of a few hundred meV. This height is to be compared with the thermal energies of carriers
at room temperature ($\approx 26$\,meV). Hence, the thermal motion of the electrons does not allow them to frequently cross
the barriers, and quantum mechanical tunneling is the main transport phenomena at length scales less than the mean free
path of the electrons.

From Fig.~\ref{fig:heterostructure}(i) it is evident that the conduction band potential step is given by $\Delta E_c = \chi_1 - \chi_2$. Because the Fermi level must be continuous in chemical and thermal equilibrium, the valence band potential step is found to be $\Delta E_v = E_{G2} - E_{G1}-\Delta E_c$.

The I-V characteristics of a one-dimensional quantum heterostructure can be related to the transmission probability according to the relation\cite{tsuesaki}
\be
\label{eq:IV}
J = \frac{e}{4\pi^3 \hbar}\!\int_0^\infty dk_l \!\int_0^\infty dk_t [f(E)-f(E')] T^* T \frac{\partial E}{\partial k},
\ee
where $k_l$ and $k_t$ are the wave vectors in the longitudinal and transverse directions respectively, and $f$ is the density of states given by the Fermi-Dirac distribution.
The theoretical results can be compared with experimental data available from the current-voltage characteristics of the device. The mass to be used in the Schr\"{o}dinger equation is the effective mass of the electron $m*$ in the longitudinal direction. As mentioned in Ref.~[23] 
the electrons lose coherence after tunneling through a distance of the order of the mean free path of the carriers, resulting in a widening of the peaks in the I-V characteristics. Calculations of the bound and quasi-bound states
of quantum heterostructures are available.\cite{ieee}

There are two major approximations in the modeling of semiconductor heterostructures by rectangular potentials. The effective mass $m*$ of the carrier changes when the electrons pass through a heterojunction, but this variation is not incorporated in the analysis. Also, the potential profile is not steplike -- the energy bands of the two materials at the heterojunction change smoothly because of factors such as the inhomogeneities present at the interface, space-dependent composition of the compounds, and possible mechanical strain of the layers. The best known example is the SiO$_2$/Si heterojunction with a very small density of defects at the interface. Compound semiconductors usually have a larger concentration of defects and inhomogeneities, leading to a continuous barrier rather than a rectangular one.\cite{chandkumar, zeyrek} The experimental data indicate a double Gaussian distribution of heights for the potential barriers. Another major reason for the smoothening of the heterojunction potential shapes is the formation of a space charge.\cite{chebotarev}

Our results may be used to compare models with experimental
data which are available for semiconductor heterostructures. This aspect requires more work -- we need to find out the theoretical
I-V characteristics from the transmission coefficients obtained by
using Eq.~\eqref{eq:IV}. In principle,
other functional forms may also be chosen and the I-V characteristics
obtained and compared with experiments to determine the
actual distribution of barrier heights. It is possible that such
investigations may help design device applications where precise knowledge
about fabricated structures is often very useful.

\section{REMARKS}

We have carried out similar computations for the Lorentz barrier, and obtained results closer to the rectangular barrier than to the Gaussian barrier. Other functional forms of the barriers may be studied to learn their
characteristic features. A further challenging problem
is to consider multiple barriers and external, applied field effects, which
are necessary to model realistic
systems in condensed matter/semiconductor physics.

\begin{acknowledgements}
We express our gratitude to the anonymous reviewers for their valuable suggestions in improving
the article. \textcolor{black}{We would also like to thank Prof. Sudipta Sarkar and Dr. Kabir Chakravarti for pointing out a misplaced parenthesis in Eq.~\eqref{eq:sqtrans}}. 
\end{acknowledgements}

\end{document}